\newcommand{\w}{\omega}

\newcommand{\TN}{T_{\rm N}}
\newcommand{\Tf}{T_{\rm f}}
\newcommand{\TCW}{\Theta_{\rm CW}}

\newcommand{\Msat}{M_{\rm sat}}

\newcommand{\op}{\varphi} 

\newcommand{\Ztwo}{$\mathbb{Z}_2$}

\newcommand{\SUtwo}{SU(2)}
\newcommand{\SOfive}{SO(5)}
\newcommand{\Uone}{U(1)}
\newcommand{\ON}{O($N$)}

\newcommand{\priro}{Pr$_2$Ir$_2$O$_7$}
\newcommand{\irhyp}{Na$_4$Ir$_3$O$_8$}
\newcommand{\bacusio}{BaCu$_2$SiO$_6$}
\newcommand{\yrs}{YbRh$_2$Si$_2$}
\newcommand{\ceau}{CeCu$_{6-x}$Au$_x$}
\newcommand{\ceag}{CeCu$_{6-x}$Ag$_x$}
\newcommand{\ET}{BEDT-TTF}
\newcommand{\kaET}{$\kappa$-(ET)$_2$}
\newcommand{\kaETCN}{$\kappa$-(ET)$_2$Cu$_2$(CN)$_3$}
\newcommand{\rucl}{$\alpha$-RuCl$_3$}
\newcommand{\ybal}{$\beta$-YbAlB$_4$}

\newcommand{\eqref}[1]{{(\ref{#1})}}

\documentclass[10pt]{iopart}
\usepackage{iopams} 
\usepackage{tabularx,graphicx}
\usepackage{color}

\begin{document}

\title[Frustration and quantum criticality]{Frustration and quantum criticality}

\author{Matthias Vojta}
\address{Institut f\"ur Theoretische Physik, Technische Universit\"at Dresden, 01062 Dresden, Germany}

\begin{abstract}
This review article is devoted to the interplay between frustrated magnetism and quantum critical phenomena, covering both theoretical concepts and ideas as well as recent experimental developments in correlated-electron materials. The first part deals with local-moment magnetism in Mott insulators and the second part with frustration in metallic systems. In both cases, frustration can either induce exotic phases accompanied by exotic quantum critical points or lead to conventional ordering with unconventional crossover phenomena. In addition, the competition of multiple phases inherent to frustrated systems can lead to multi-criticality.
\end{abstract}

\ioptwocol
\tableofcontents
\markboth{Frustration and quantum criticality}{Frustration and quantum criticality}


\section{Introduction}
\label{sec:intro}

A physical system is commonly called frustrated if not all contributions to its potential energy can be simultaneously minimized. This happens frequently for systems of magnetic moments, namely if the minimization of all interaction energies poses incompatible constraints on the system's configuration. The perhaps simplest example is given by antiferromagnetically coupled Ising spins on a triangle.
Frustration can arise from the geometry of the underlying lattice and/or from the nature of the interactions. The most obvious effect of frustration is to counteract the usual tendency towards symmetry-breaking order at low temperatures. As a result, a frustrated system may either have a strongly reduced ordering temperature or show no order at all, the latter often leading to exotic liquid-like phases. In addition, the suppression of conventional ordering phenomena can induce a competition of multiple less conventional phases, resulting in complex phase diagrams, non-trivial crossover phenomena, an accumulation of entropy at low temperature, and a large sensitivity to tuning parameters.

The past decade has seen a flurry of interest in frustrated systems \cite{ramirez94,balents_nat10,starykh13,savary_rop17,kanoda_rmp17,lacroix_book,diep_book}, primarily driven by the search for novel states of matter. Prime examples are spin liquids with fractionalized degrees of freedom, skyrmion lattices with emergent artificial electrodynamics, fractionalized Fermi liquids, and their descendants. Many of these phases are characterized by non-trivial topological properties. As a result, phase transitions in and out of these phases often do not follow the conventional paradigms of Landau, rendering the study of quantum phase transitions in frustrated system a fascinating subject.

The purpose of this article is to review conceptual aspects and recent developments concerning quantum phase transition in systems with frustration, with the focus on systems of interacting electrons in solids. As will become clear below, many aspects of this young field are far from being understood, and we will highlight directions of future research.
The focus will be on thermodynamic and linear-response properties; phase transitions far from equilibrium constitute a separate interesting topic which is beyond the scope of this article. Most of the discussion will be restricted to systems in spatial dimensions $d\geq 2$, mainly because the case $d=1$ is special in many respects; for a detailed discussion on one-dimensional (1D) correlated systems we refer the reader to Ref.~\cite{giamarchi_book}.
We finally note that this article will not cover transitions involving topological states of band electrons; those have become a very rich field in itself and are reviewed e.g. in Ref.~\cite{qi_rmp11}.

This article will discuss theoretical ideas and concrete results -- both analytical and numerical -- for particular models, and will also make contact with relevant experimental data. In the context of numerical simulations it must be kept in mind that the available tools for frustrated quantum systems in $d=2,3$ are rather limited, as the so-called sign problem often prohibits efficient numerical simulations using Quantum Monte Carlo (QMC) techniques, and other methods often suffer from serious finite-size limitations. Hence, for many models reliable numerical results at low temperature, which would allow one to extract critical properties, are not available.

\subsection{Outline}

The body of this article is organized as follows:
It starts with a micro-review of quantum criticality in Sec.~\ref{sec:qcprimer}, followed by two sections on magnetic insulators:
Sec.~\ref{sec:fruins} summarizes the main theoretical concepts of frustrated magnetism in insulators, including spin liquids, valence-bond solids, and order-by-disorder phenomena.
Sec.~\ref{sec:critins} then turns to the interplay of frustration and criticality, by discussing theoretical ideas and results for quantum critical phenomena in frustrated magnetic insulators.
We then turn to metals: Sec.~\ref{sec:frumet} explains various ways of how frustration can enter metallic systems; this covers frustrated Hubbard models as well as Kondo lattices and other multiband systems.
This is followed by Sec.~\ref{sec:critmet} which reviews theories for quantum phase transitions in frustrated metals.
Further experimentally relevant ingredients as discussed in Sec.~\ref{sec:ingredients}, such as broad crossover regimes, multi-criticality, and the influence of quenched disorder, the latter often leading to glassy behavior.
The final section~\ref{sec:exp} is devoted to a brief discussion of experimental results on selected materials, aimed at connecting theory and experiment.
A discussion of open questions closes the article.


\section{Quantum criticality primer}
\label{sec:qcprimer}

This is not a review article on quantum criticality in general, and we refer the reader to Refs.~\cite{ssbook,mv_rop} for a detailed exposure. Here we summarize a few important aspects in a nutshell.

\begin{figure}
\center
\includegraphics[width=0.8\linewidth]{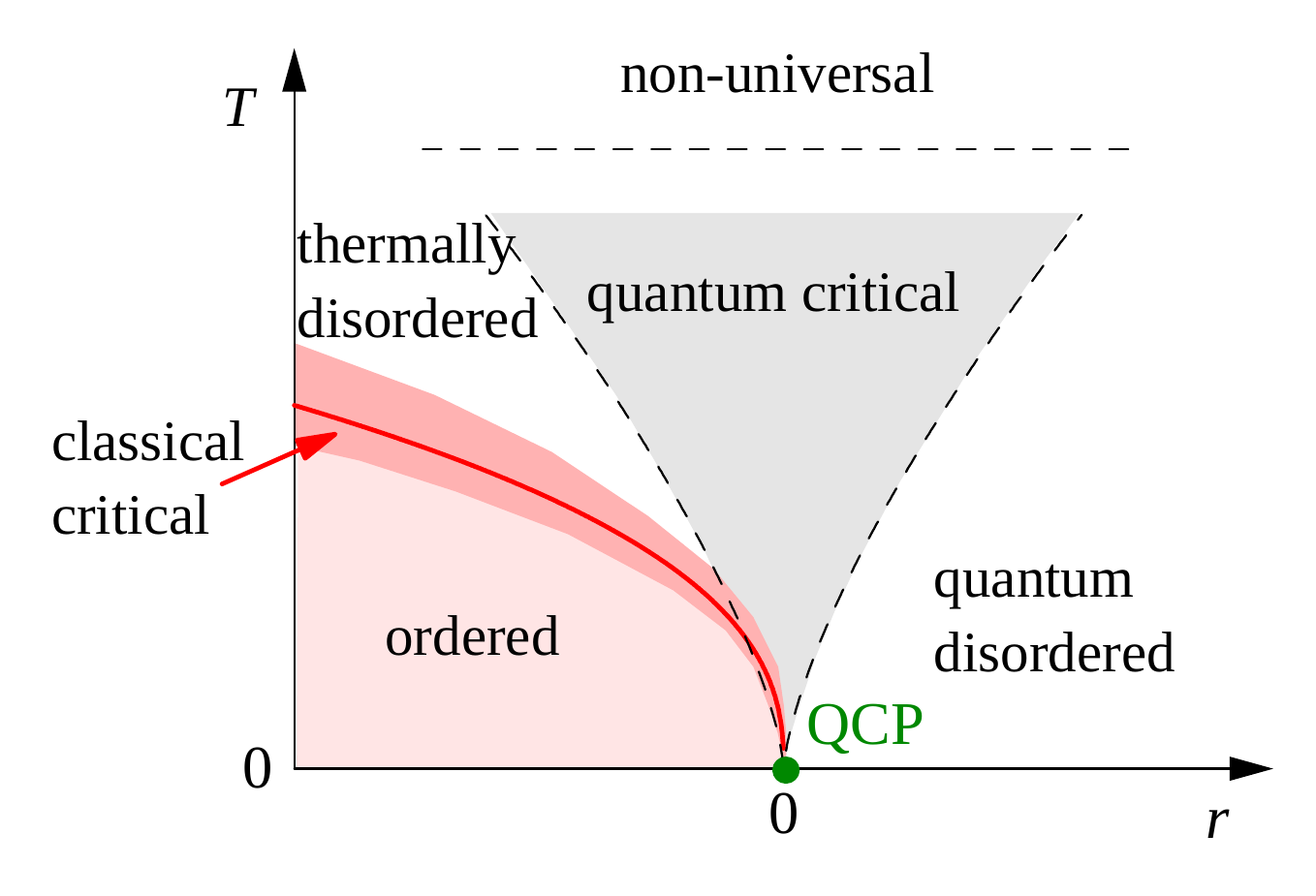}
\caption{
Generic phase diagram in the vicinity of a quantum critical point as function of a non-thermal control parameter $r$ and temperature $T$. An ordered phase exists for $r<0$ and low $T$, bounded by a line of classical phase transitions which terminates at the QCP at $r=0$, $T=0$. The quantum critical regime is defined by $k_B T \gg |r|^{\nu z}$, where $\nu$ and $z$ are the correlation length and dynamical exponents.
}
\label{fig:pd}
\end{figure}

A quantum phase transition (QPT) is a phase transition taking place at temperature $T\!=\!0$ upon tuning a non-thermal control parameter like pressure or magnetic field. The finite-temperature properties near a continuous QPT are highly unusual: Due to the peculiar properties of the quantum ground state at the transition point, dubbed quantum critical point (QCP), the so-called quantum critical regime located at finite $T$ above the QCP, Fig.~\ref{fig:pd}, displays properties distinct from that of any stable phase of matter. These properties include power-law behavior with unconventional exponents of thermodynamic and transport quantities as function of absolute temperature as well as scaling behavior, where suitably rescaled observables depend only on dimensionless ratios of external parameters.

From a theoretical perspective, the universal properties of QPTs can often be described using a continuum quantum field theory for the transition's order parameter, where the choice of the latter is dictated by the way in which symmetries of the Hamiltonian are spontaneously broken at the transition. This goes back to ideas of Landau who pioneered the ideas of symmetry breaking and local order parameters in the context of phase transitions. This concept was later extended to quantum phase transitions by taking into account fluctuations of the order parameter in imaginary time, i.e., quantum fluctuations -- this leads to the so-called Landau-Ginzburg-Wilson (LGW) approach. Importantly, the quantum dynamics of the order parameter can be non-trivial due to damping or Berry-phase terms.

For Mott-insulating quantum magnets the LGW theory for a zero-temperature transition between a featureless paramagnet and, e.g., a collinear ordered antiferromagnet takes the form of a quantum $\op^4$ model with the action
\begin{equation}
\label{phi4qu}
  S = \int d^{d}x \int_{0}^{\beta}\!d\tau \left( \frac{c_0^{2}}{2}(\partial_i \vec{\op})^{2}
    + \frac{1}{2}(\partial_\tau \vec{\op})^{2} + \frac{\delta_0}{2}\vec{\op}^{2} + \frac{u_0}{4!}(\vec{\op}^2)^2 \right)
\end{equation}
where $\partial_i=\partial/\partial x_i$, and $\vec{\op}(\vec{x},\tau)$ is a local $N$-component order-parameter field which is assumed to vary slowly in space and time and encodes the ordering tendency at a microscopic wavevector $\vec{Q}$.
Further, $\tau$ is imaginary time, and $c_0$, $\delta_0$, and $u_0$ are parameters. Decreasing the non-thermal control parameter $\delta_0$ at low temperature tunes a transition between a disordered and an ordered phase, with the {\ON} symmetry spontaneously broken in the latter; $N=3$ for magnetic ordering in the presence of {\SUtwo} spin symmetry. More precisely, $\delta_0$ acquires a temperature-dependent renormalization, and the transition occurs at $\delta_0=\delta_c$ where the renormalized $\delta$ vanishes. The distance to the QCP can be expressed as
\begin{equation}
r = \delta_0-\delta_c(T=0)
\end{equation}
and may be tuned by pressure or chemical composition.

In Eq.~\eqref{phi4qu} space and time enter symmetrically, corresponding to a dynamical exponent $z=1$. The time direction in the integral may be interpreted as an additional space direction, such that the quantum theory in $d$ dimensions at $T\!=\!0$ is equivalent to a classical theory in $D=d+z$ dimensions. While the local order-parameter description with $z=1$ applies to many QPT in insulators, the situation in metals is more complicated due to the presence of low-energy fermionic excitations.
Two additional remarks are in order:
(i) QPTs into ferromagnetic or polarized phases in the presence of {\SUtwo} spin symmetry follow a quantum dynamics different from that of the $\op^4$ model because a \emph{conserved} density changes across the transition, see Sec.~\ref{sec:bec}.
(ii) Berry-phase terms, which are generically present in a field-theory description of spin systems, do not appear in Eq.~\eqref{phi4qu} because they are irrelevant for the transition between featureless paramagnet and antiferromagnet. They are, however, responsible for much of the physics beyond LGW which will be described in Sec.~\ref{sec:critins}.

For finite-temperature (i.e. classical) transitions, the upper critical dimension above which mean-field critical behavior is realized is $D_c^+=4$ for a standard $\op^4$ theory. In the quantum case, the presence of temporal fluctuations implies that the upper critical dimension for QPTs is given by $d_c^+=4-z$. For instance, continuous QPTs in $d=3$ with $z=1$ display mean-field behavior with logarithmic corrections. For phase transitions involving fermions the situation may be more complicated, see Sec.~\ref{sec:critmet}.

In the phase diagram in Fig.~\ref{fig:pd}, the system is critical at the QCP, located at $r=0$ and $T=0$ (and possibly also along the classical transition line at finite $T$). The quantum critical regime corresponds to the universal high-temperature regime of the field theory \eqref{phi4qu}, and is defined by $k_BT\gg|r|^{\nu z}$ with $\nu$ being the correlation-length exponent.
The QCP can be approached either by varying the tuning parameter $r$ at $T=0$ or by lowering $T$ at fixed $r=0$. Interestingly, the singular behavior of thermodynamic quantities for these two cases can be related: A suitable diagnostic is the so-called Gr\"uneisen parameter, in the following explicitly formulated for pressure taking the role of $r$. The Gr\"uneisen parameter is defined as the ratio between thermal expansion $\alpha=(1/V) (\partial V/\partial T)_p$ and specific heat $C_p=T (\partial S/\partial T)_p$,
\begin{equation}
\label{gammadef}
\Gamma=\frac{\alpha}{C_p} =
-\frac{1}{V_m T}\frac{(\partial S/\partial p)_T}{(\partial S/\partial T)_p}
\end{equation}
where $V_m=V/N$ the molar volume. If one takes the ratio of the singular parts of $\alpha$ and $C_p$, the scaling dimensions of $T$ and $S$ cancel, such that a {\em universal} divergence in the low-$T$ limit emerges \cite{markus}
\begin{eqnarray}
\Gamma_{\rm cr}(T=0,r)&=&\frac{\alpha_{\rm cr}}{C_{p,\rm cr}} = G_r |r|^{-1}\,,
\label{gr1}
\\
\Gamma_{\rm cr}(T,r=0)&=& G_T T^{-1/(\nu z)}\,.
\label{gr2}
\end{eqnarray}
$G_r$ and $G_T$ are prefactors, where $G_r$ is universally given by a combination of critical exponents, $G_r=\nu(d-z)$, for details see Ref.~\cite{markus}. Given that $\Gamma$ remains finite at a finite-$T$ phase transition, a divergence of $\Gamma$, accompanied by a sign change as function of $r$ at low $T$, is considered a unique signature of a continuous QPT.
For a transition driven by magnetic field $H$ instead of pressure, the suitable analogue is the magnetic Gr\"uneisen parameter,
\begin{equation}
\label{gammahdef}
\Gamma_H=-\frac{dM/dT}{C} =
\frac{1}{T}\left(\frac{\partial T}{\partial H}\right)_S
\end{equation}
where $M$ is the magnetization. The second equality shows that $\Gamma_H$ measures the adiabatic magnetocaloric effect.


\section{Frustrated magnetism in local-moment systems}
\label{sec:fruins}

To set the stage, we introduce important concepts for frustrated magnets. In this section, we will cover the physics of Mott insulators with local moments; aspects of frustrated metals will be discussed in Sec.~\ref{sec:frumet}.
We will consider systems of local moments, i.e., quantum-mechanical spins transforming as {\SUtwo} vectors, placed on the sites of a regular lattice, with a Hamiltonian containing two-spin interactions plus, perhaps, multi-spin exchange terms. The most generic model Hamiltonian is an antiferromagnetic Heisenberg model with nearest-neighbor interactions,
\begin{equation}
\label{hheis}
\mathcal{H} = J \sum_{\langle ij\rangle} \vec{S}_i \cdot \vec{S}_j
\end{equation}
where the $\vec{S}_i$ represent spins of given size $S$ located on sites $i$, and $J>0$ is the exchange interaction. We will also discuss spin-anisotropic interactions (e.g. of Ising type), arising due to spin-orbit coupling.

The Heisenberg interaction in Eq.~\eqref{hheis} favors antiparallel moments on neighboring lattice sites. Consequently, this interaction is non-frustrated on lattices where all closed loops of interaction paths have even length, such that an alternating up--down arrangement, corresponding to collinear magnetic order, can cover the lattice. This applies to the square and cubic lattices as well as, e.g., the honeycomb lattice.
In contrast, frustration is induced on lattices with odd-length loops, e.g. the triangular, kagome, bcc, fcc, and pyrochlore lattices. On some of these lattices, a magnetically ordered ground state -- often non-collinear -- is realized for any $S$ despite the existence of frustration, the triangular lattice with its $120^\circ$ order being an established example, while in other cases order may be entirely absent.

In addition to the described {\em geometric frustration}, rooted in the geometry of the underlying lattice, incompatible constraints may be caused by the nature of the exchange interactions, leading to {\em exchange frustration}. A prominent case are so-called Kitaev interactions \cite{kitaev06} (and the related compass interactions \cite{compass_rmp}), where different spin components interact along different bond directions. Such a situation may arise from strong spin-orbit coupling and has been shown to induce a variety of non-trivial magnetic phases.

Given that frustration tends to suppress magnetic order, a popular experimental way to quantify frustration in a given system is the so-called frustration ratio, $f=|\TCW|/\TN$, where $\TN$ is the ordering temperature and $\TCW$ the Curie-Weiss temperature, the latter being a measure for the strength of exchange interactions \cite{ramirez94}. Materials with $f>5$ are commonly called ``frustrated''. The extreme case of no long-range order (LRO) down to $T\!=\!0$, formally $f=\infty$, then corresponds to a ground state with only short-range correlations. A regime with highly correlated but fluctuating spins and no LRO at temperatures $T\ll|\TCW|$ is often dubbed ``spin liquid'' (although more precise definitions are available, see below).

\subsection{Classical spin liquids}

In the classical limit, formally obtained for spin size $S\to\infty$, spins can be viewed as unit vectors, and non-trivial commutators vanish. In the presence of frustration, not all Hamiltonian terms can be simultaneously minimized.
This may lead to a classical ground state which is either unique up to global symmetry transformations -- in this case the system is called ``weakly frustrated'' -- or which has degeneracies scaling with the system size, rendering the system ``strongly frustrated''.\footnote{Intermediate cases with sub-extensive degeneracies exist as well.} In the latter case, the resulting manifold of lowest-energy states defines a \emph{classical spin liquid}.
A celebrated example is spin ice, referring to moments with local Ising anisotropy and ferromagnetic interactions on a pyrochlore lattice, viz. a lattice of corner-sharing tetrahedra \cite{spinice}.

Often, a classical spin liquid can be characterized by a set of local conditions which define the ground-state manifold (but \emph{not} a unique state up to global symmetry transformations, as explained above). Examples are the conditions ``two in, two out'' for the Ising configurations of individual tetrahedra of spin ice or the condition $\sum_\triangle \vec{S}_i = 0$ for the spin configurations of a kagome-lattice Heisenberg model.
Hence, these conditions {\em underconstrain} the manifold of states; recall that the original problem of minimizing all Hamiltonian terms simultaneously {\em overconstrains} the manifold of states if frustration is present.
Local constraints can often formulated as an emergent lattice gauge theory. For instance, the ``two in, two out'' condition can be translated into ${\rm div}\,b =0$ where $b$ is an artificial magnetic field and $\rm div$ a suitably defined lattice divergence.

For Ising spins (i.e. with countable number of states) a classical spin liquid can be characterized by an extensive ground-state entropy $S_0/N$ where $N$ is the number of lattice sites. Typical examples are the Ising model on a triangular lattice, with $S_0/(N k_B) \approx 0.323$ \cite{wannier50}, and classical spin ice, with $S_0/(N k_B) \approx 1/2 \ln(3/2) \approx 0.203$ \cite{pauling}.
For classical spin liquids made from XY or Heisenberg spins a residual entropy cannot be defined, but the degeneracy may be quantified via the difference between the number of continuous degrees of freedom and the number of local constraints.

Elementary excitations of classical spin liquids correspond to configurations which violate one (or more) of the local ground-state conditions; in the gauge-theory language these become elementary charges. For spin ice, the excitations are tetrahedra with ``three-in, one-out'' or ``one-in, three-out'' configurations; these have been shown to behave like magnetic monopoles upon including dipolar interactions \cite{cms08}.

\subsection{Quantum spin liquids}
\label{sec:qsl}

With quantum fluctuations included, frustrated systems may realize local-moment states without symmetry breaking and only short-range order down to lowest temperatures. Such quantum spin liquids (QSLs) \cite{balents_nat10,savary_rop17,kanoda_rmp17} display some differences compared to their classical counterparts:
(i) Quantum fluctuations typically remove the extensive ground-state degeneracy of strongly frustrated systems by quantum tunnelling, resulting in unique ground states (up to global symmetry transformations or topological degeneracies).
(ii) QSLs are thermodynamically stable phases of matter, characterized by emergent dynamic gauge fields and topological order. This implies the existence of fractionalized excitations which are coupled to the gauge field. Despite this coupling, the fractionalized excitations are asymptotically free, i.e., deconfined.
(iii) The wavefunctions of QSLs can be characterized by long-range entanglement \cite{a1KP06,a1LW06}.
Importantly, QSLs need to be distinguished from ``trivial'' quantum paramagnets without topological order and fractionalization. Examples of the latter are coupled-dimer magnets where pairs of spins 1/2 form magnetic singlets, as realized e.g. in TlCuCl$_3$.

Different types of QSLs can be distinguished depending on the spectrum and statistics of the emergent excitations and on the gauge structure. Prominent examples are fully gapped {\Ztwo} spin liquids, for which topological order can be sharply defined, and algebraic {\Uone} spin liquids with gapless excitations. For an in-depth discussion of topological order and attempts of classifications we refer the reader to the literature \cite{balents_nat10,savary_rop17,wen02}.
Relevant to the existence of non-trivial many-body states is a theorem due to Lieb-Schulz-Mattis \cite{LSM61} and its higher-dimensional generalization by Hastings \cite{hastings04}. It states that in a system with half-odd-integer spin per unit cell and global {\Uone} symmetry, the excitation spectrum in the thermodynamic limit cannot simultaneously fulfill the two conditions: (a) the ground state is unique and (b) there is a finite gap to all excitations. This implies that a gapped symmetry-unbroken state must have a ground-state degeneracy which is topological in nature.
We finally note that, conceptually, topological order and fractionalization may co-exist with spontaneous symmetry breaking: For instance, broken time-reversal symmetry on top of a spin liquid leads to a chiral spin liquid, while magnetic long-range order leads to a fractionalized ordered magnet.

An intuitive picture of a QSL with underlying {\SUtwo} symmetry is provided by the resonating valence-bond (RVB) idea, originally proposed by Anderson for the triangular-lattice Heisenberg model \cite{rvb}. RVB refers to pairing spins on a lattice into singlets and then forming a quantum superposition of many different pairings, i.e., different dimer coverings of the lattice, such that the symmetries of the Hamiltonian are preserved.\footnote{The first existence proof of a {\Ztwo} spin liquid was given for a triangular-lattice quantum dimer model which realizes an RVB phase \cite{ms01}.} This picture captures the aspect of fractionalized excitations, as the breaking of a dimer leads to two monomer excitations with independent dynamics: These monomers are objects carrying charge $0$ and spin $1/2$, typically called {\em spinons}. In a {\Ztwo} spin liquid, they are coupled to an emergent {\Ztwo} gauge field, whose excitations are {\Ztwo} vortices (or fluxes) called {\em visons.}

A well-studied spin model with geometric frustration is the Heisenberg model on the kagome lattice. For quantum spins $1/2$, with antiferromagnetic interactions as in Eq.~\eqref{hheis}, there is strong numerical evidence that this realizes a fractionalized QSL. However, the nature of this QSL has not been conclusively clarified to date, as numerical results have been interpreted in favor of either a gapped {\Ztwo} spin liquid \cite{white11} or a {\Uone} spin liquid with a Dirac-cone spectrum \cite{iqbal13,pollmann17}. A candidate material realizing the kagome-lattice spin-$1/2$ Heisenberg model is Herbertsmithite, ZnCu$_3$(OH)$_6$Cl$_2$, which indeed displays spin-liquid-like behavior \cite{helton07,norman_rmp16}. However, the role of quenched disorder is debated \cite{norman_rmp16}.
Numerical evidence for QSL phases in Heisenberg models of spins $1/2$ has also been found for square \cite{balents11} and triangular-lattice models \cite{white15} with first and second-neighbor interaction, so-called $J_1$-$J_2$ models.

Spin systems without {\SUtwo} spin symmetry have an even richer phenomenology. A by now popular route to QSLs was proposed by Kitaev \cite{kitaev06}: A model with bond-dependent Ising interactions on a honeycomb lattice realizes an exactly solvable {\Ztwo} spin liquid whose emergent excitations are Majorana fermions and static {\Ztwo} gauge fluxes. This model has been subsequently generalized to other lattices and dimensions \cite{hermanns16}.
Experimentally, strong Kitaev interactions on a honeycomb lattice have been deduced for the materials {\rucl} \cite{plumb14,banerjee17}, Na$_2$IrO$_3$ \cite{sin10,chun15}, and various polytypes of Li$_2$IrO$_3$ \cite{singh2012,takayama15,modic14,kimchi15}; however, all of these materials display magnetic LRO at low temperatures due to the presence of additional interactions.

\subsection{Valence-bond solids}

An alternative quantum paramagnetic state of spins $1/2$ that can be constructed from dimer coverings of the underlying lattice is a so-called valence-bond solid. In this state, the wavefunction is dominated by a single covering with a periodic arrangement of dimers. As a result, the state spontaneously breaks translation and rotation symmetry of the lattice, hence the label {\em solid}. Excitations of valence-bond solids carry integer spin, i.e., spinons are confined.

Variants of valence-bond solids can be constructed for larger constituent spins and/or from larger units, the common theme being that the state in the resulting magnetic unit cell represents a spin singlet. For instance, plaquette valence-bond solids with unit cells of four spins $1/2$ have been discussed for the square-lattice checkerboard and $J_1$-$J_2$ models.

\subsection{Order by disorder and unconventional types of order}
\label{sec:obd}

In addition to phases with unbroken spin symmetry, like spin liquids and valence-bond solids,  frustrated spin systems can of course display phases with broken spin symmetry, both conventional and unconventional \cite{lacroix_book}.

First, conventional magnetic order can emerge in an unconventional way. Most prominent is so-called ``order by disorder'' which refers to a situation where a frustration-induced degeneracy of the classical ground-state manifold is lifted by fluctuations, either thermal or quantum \cite{villain80}. A well-studied example is the easy-plane pyrochlore antiferromagnet, where long-range order emerges due to fluctuations from a one-parameter manifold of classically degenerate states \cite{zhito12}.

Second, less conventional magnetic order can appear as a result of large crystallographic unit cells or non-Heisenberg interactions. Among the possibilities are so-called multi-$Q$ states where the ordering pattern results from the superposition of modulations with multiple inequivalent wavevectors, among which skyrmion lattices have attracted particular attention \cite{rbp06}.

Third, ordered states may spontaneously break spin symmetry not by dipolar order, but by order in higher multipole channels. The simplest form is quadrupolar or spin-nematic order which breaks {\SUtwo} symmetry and is described by a local rank-2 tensor order parameter \cite{blume69,levy71}. Such order is known to be realized in certain spin-$1$ Heisenberg models with additional biquadratic interactions \cite{penc06}.


\section{Quantum criticality in frustrated insulating magnets}
\label{sec:critins}

As explained in Sec.~\ref{sec:qcprimer}, the standard paradigm for a continuous QPT between a featureless disordered phase and a symmetry-broken ordered phase is that of Landau-Ginzburg-Wilson where critical properties are determined by a low-energy theory of a bosonic order parameter alone.
While this simple paradigm applies for instance to ordering transitions in many coupled-dimer magnets, the situation in frustrated systems can be different for the following reasons:
(i) If a quantum paramagnetic phase is a fractionalized spin liquid, it is {\em not} featureless, because it is characterized by topological order.
(ii) The ordered-state manifold may be unconventional, i.e., not be characterized by a local order parameter or by a unique ordering wavevector. Long-range order may arise exclusively from fluctuation effects.
(iii) A transition might occur between states without spontaneous symmetry breaking.
(iv) The active quantum degrees of freedom can be different from the fluctuations of the order parameter, i.e., if a local order parameter exists, it might be a composite when expressed in the elementary degrees of freedom.
(v) Frustration may enhance fluctuations such that the transition is rendered first order.

In the following we will list a few aspects of and concrete proposals for critical theories in frustrated magnetic insulators; some of the general remarks apply both to classical and quantum phase transitions. In fact, many of the non-trivial quantum theories are formulated in fractionalized degrees of freedom.

\subsection{Conventional ordering transitions}
\label{sec:convqpt_ins}

The simplest case, a quantum transition from a featureless paramagnet to a symmetry-broken phase with antiferromagnetic or VBS order, is expected to be described by an LGW theory of $\op^4$ type, Eq.~\eqref{phi4qu}, with dynamical exponent $z=1$. Symmetry and wavevector of the order parameter determine the effective number of order-parameter components and the structure of the interaction terms in the field theory.

Frustration enters in a non-trivial way, because the order-parameter structure of non-collinear or non-coplanar states is much richer than that of simple collinear magnets. Most straightforwardly, this translates into a larger number of components $N$ in the corresponding $\op^4$ theory. This is not all: For instance, a non-collinear ordered state often breaks both {\SUtwo} spin rotation symmetry and a {\Ztwo} chiral symmetry, and both symmetries can be broken either in a single or in two separate transitions. For the classical case, this has been studied for stacked triangular-lattice Heisenberg antiferomagnets: Monte Carlo simulations have observed a single transition with non-trivial critical exponents, different from that of standard {\ON} universality, consistent with a proposal by Kawamura \cite{kawamura85}.\footnote{More recent theory works predict the transition in stacked triangular-lattice Heisenberg antiferromagnets to be weakly first order \cite{schotte88,tissier00}.} Numerical results for the quantum case are, to our knowledge, not available due to the notorious sign problem.

More seriously, frustration can render invalid the concept of discrete well-defined wavevector for critical fluctuations: Upon approaching an ordered state, fluctuations may become soft on a manifold of wavevectors, e.g., owing to frustration-induced degeneracies. Strong fluctuation effects may then cause the transition to be first order, see Sec.~\ref{sec:brazov}. Alternatively, exotic novel intermediate phases might emerge.
An interesting open problem in this context constitutes the quantum melting of a skyrmion crystal \cite{rbp06}. Such a phase has been observed in a number of helical magnets in the presence of a magnetic field, e.g., in MnSi \cite{pfleiderer09}. In MnSi, long-range magnetic order can be suppressed by the application of pressure, giving way to an extended non-Fermi liquid phase at low temperature \cite{mnsi_nfl}. It has been speculated that this behavior is related to partial order, e.g., a skyrmion liquid, but a concise theory is not known.

A further complication, frequently present in strongly frustrated systems, arises due to order-by-disorder physics (Sec.~\ref{sec:obd}): If the actual ordered state is selected by fluctuation effects from a larger (e.g. classically degenerate) manifold, then some or all properties of the transition may be determined by the larger symmetry of this manifold.
This type of physics is known from $Z_n$ clock models, or alternatively XY models with $Z_n$ anisotropy. Here, anisotropies with $n\geq n_c$ are irrelevant at criticality, such that the critical behavior is that of the XY model. For $d=2$ (or $D=1+1$) this even changes the phase diagram, as an intermediate critical phase intervenes between the disordered and the $Z_n$-ordered phases for $n\geq 5$ \cite{jose77}. An example of recent interest are the finite-temperature intermediate phases present in the two-dimensional (2D) Heisenberg-Kitaev model \cite{cjk10} where the relevant ordered phases are sixfold degenerate as a result of Kitaev interactions reflecting spin-orbit coupling \cite{perkins13}. Theoretical results for the \emph{quantum} phase transitions in this model indicate first-order behavior both on analytical \cite{schaffer12} and numerical \cite{cjk13,gohlke17} grounds, but the numerics has not reached conclusive accuracy yet.

Strong frustration may, in addition, lead to {\em dimensional reduction}: This refers to a situation where the effective spatial dimension of the order-parameter fluctuations is smaller than that expected from the microscopic model. For instance, a three-dimensional (3D) layered system with inter-layer frustration may display 2D critical behavior. Experimentally this applies, e.g., to \bacusio, see Sec.~\ref{sec:bacusio} for a more detailed discussion.\footnote{An interesting instance of dimensional reduction has been recently reported for the field-driven quantum phase transition in LiErF$_4$, an dipolar XY-type antiferromagnet \cite{ronnow12}. Its origin is likely in the frustrated nature of the dipolar interaction, but not fully understood.} Such dimensional reduction typically does not reach down to lowest energies and temperatures, due to residual higher-dimensional couplings, such that a dimensional crossover to fully 3D critical behavior at lowest temperatures occurs \cite{garst08}.
Parenthetically, we note that dimensional reduction can also arise from the interplay of frustrated lattice and orbital structures: For instance, $t_{2g}$ orbitals placed on the B-sites of a spinel lattice have a strong direct overlap along 1D chains, leading non-trivial ordered states, e.g., in MgTi$_2$O$_4$, CuIr$_2$S$_4$ and AlV$_2$O$_4$ \cite{khomskii11}.

\subsection{Transitions involving spin-liquid states}
\label{sec:sltrans}

QPTs in and out of topological spin-liquid states are fundamentally different from the conventional transitions discussed above, as they necessarily involve the fractionalized degrees of freedom of the spin liquid. In many cases, these are spinons (i.e. fractionalized constituents of the microscopic spins) and excitations of the emergent gauge field in its deconfined phase. Continuous transitions out of a spin liquid can often be understood as a condensation transition of one of these particles (or bound states thereof) \cite{savary_rop17,slinger09}. Physical spins are then composite objects in terms of the critical degrees of freedom. As a result, spin correlation functions display critical power laws with {\em large} anomalous exponents: While standard {\ON} universality yields numerically small anomalous exponents, e.g. $\eta=0.06$ for the 3D Heisenberg model, many of the exotic transitions discussed below have $\eta$ values for physical correlators of order unity.

Starting from a fractionalized spin liquid, one can envision the following options for QPTs:
(i) a confinement transition to a featureless paramagnet,
(ii) a confinement transition with concomitant symmetry breaking, leading to e.g. magnetic or VBS order -- typically these are Higgs-type transitions driven by the condensation of a particle with gauge charge,
(iii) a condensation transition which leaves the deconfinement intact, which then leads to exotic fractionalized magnetic (AF$^\ast$) or VBS states (VBS$^\ast$),
(iv) a transition to a different fractionalized spin liquid.

In the following, we list a few examples from the theory literature. The field theories are typically written down in terms of fractionalized particles coupled to gauge fields; in some cases topological quantum field theories (most importantly, Chern-Simons theories) have also proven useful. Most considerations apply to two space dimensions; less work has been done for $d=3$.

Transitions in group (i) require the presence of a featureless paramagnetic phase in addition to a topological spin liquid: The former can be realized, e.g., by application of a magnetic field or by the formation of singlet dimers as in bilayer models.
A concrete example is the 2D toric-code model \cite{toric} in a longitudinal field \cite{kps09}: It displays a continuous transition from a {\Ztwo} topological spin liquid to a featureless high-field phase. The transition has been shown to be in the Ising$^\ast$ universality class in $D=2+1$ dimensions \cite{schuler16}. Here, Ising$^\ast$ refers the fact that the critical degrees of freedom have Ising symmetry, but are very different from a conventional order parameter, as they derive from the fractionalized excitations of the spin liquid. Hence, thermodynamic properties are that of Ising criticality in $D=2+1$, but correlation functions of physical spins strongly differ from the conventional case as spins are composite objects here. This can be expected to generically apply to confinement transitions of {\Ztwo} spin liquids.
A second example is the ferromagnetic honeycomb-lattice Kitaev model in a magnetic field \cite{trebst11}: This displays a single transition between a {\Ztwo} spin liquid and a featureless high-field phase as well. However, it is open whether this transition is weakly first order or continuous. A third example is the pyrochlore transverse-field Ising model, which displays a first-order transition between a chiral spin liquid and a featureless high-field phase \cite{kps16}.

\begin{figure}
\center
\includegraphics[width=0.8\linewidth]{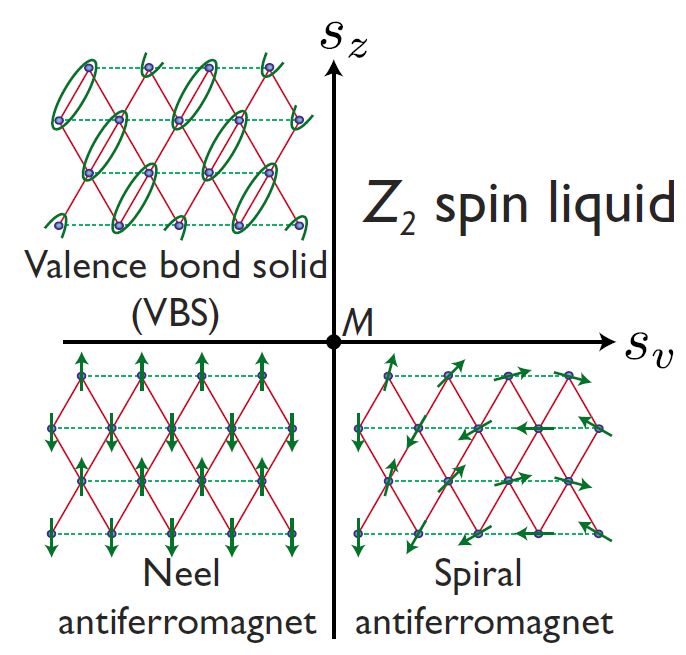}
\caption{
Global phase diagram for a 2D model of spinons with emergent {\Ztwo} gauge field, here shown for an anisotropic triangular lattice. The two parameters $s_v$ and $s_z$ represent masses of visons and spinons, respectively, in the doubled Chern-Simons theory considered in Ref.~\cite{xuss09}.
The spiral--{\Ztwo} spin liquid transition is described by a three-dimensional O(4)$^\ast$ theory, while the transition from VBS to {\Ztwo} spin liquid is of XY$^\ast$ type, see text. Further, the N\'{e}el--VBS transition is captured by a CP$^{1}$ theory (see Sec.~\ref{sec:dqcp} below), and the N\'{e}el--spiral transition is mean-field-like.
(Figure taken from Ref.~\cite{xuss09})
}
\label{fig:fourph}
\end{figure}

Transitions in group (ii) have been mainly discussed within effective field theories, and candidate models are known in many cases.
A typical situation is that of vison condensation in a 2D {\Ztwo} spin liquid; if the vison has non-trivial transformation properties under lattice symmetries, its condensation generically breaks translation symmetry and induces VBS order. Such transitions have been argued to be of \ON$^\ast$ type -- supplemented by lattice anisotropy terms which are irrelevant at criticality -- where the number of components $N$ of the vison-derived field depends on the lattice and the resulting VBS state. For example, the transition to a columnar VBS on both the square and honeycomb lattices is of 3D XY$^\ast$ type \cite{ms01,xuss09,xu11}, while on the triangular lattice the transition to a columnar VBS is proposed to be of 3D O(6)$^\ast$ type \cite{xu14}. In contrast, transitions to staggered VBS phases have been argued to be of first order \cite{xu11}. Generally, liquid--VBS transitions may be realized in Heisenberg models with further-neighbor (e.g. $J_1$-$J_2$-$J_3$) exchange interactions.
Instead of condensing visons one can consider condensing spinons in \SUtwo-symmetric {\Ztwo} spin liquids. This produces a confined antiferromagnet with spiral order via an O(4)$^\ast$ transition \cite{chub94,xuss09,moon12,sachdev16} where the symmetry arises from a doublet of complex spinon fields. A resulting ``global'' phase diagram is shown in Fig.~\ref{fig:fourph}.
Finally, condensing bound states of spinons and visons may induce conventional two-sublattice N\'eel order. At the latter transition, which is of more exotic type, both magnetic and VBS correlation functions acquire critical power laws \cite{moon12}.
In the absence of {\SUtwo} symmetry, quantum numbers need to be reconsidered, but the general picture remains valid. One example here is the 2D toric-code model perturbed by an Ising interaction which has been shown to display a continuous transition of Ising$^\ast$ type from a {\Ztwo} liquid to a ferromagnetic phase driven by defect condensation \cite{kamiya15}. A second example is the transition between a {\Ztwo} spin liquid and a superfluid phase in a Kagome-lattice XY model. This transition is in the XY$^\ast$ universality class and has been studied numerically in some detail in Ref.~\cite{isakov12}. Another case is the transition from a Kagome-lattice {\Ztwo} spin liquid to an antiferromagnet driven by Dzyaloshinskii-Moriya interactions: this is also of XY$^\ast$ type \cite{fritz10}.

A transition in group (iii) is realized upon condensing objects which do {\em not} carry gauge charge, then leading to the coexistence of symmetry-breaking order and fractionalization. For instance, condensing a gauge-neutral N\'eel vector in a spin liquid yields an AF$^\ast$ phase, and a spin-Peierls instability of a spin liquid can result in a VBS$^\ast$ state. A nice example of the latter is the instability of Majorana Fermi surfaces in 3D Kitaev-based spin liquids \cite{rosch15}.

Transitions between different spin-liquid phases, group (iv), have also been considered on the level of effective field theories. Ref.~\cite{barkeshli13} has developed a theory for transitions between chiral and {\Ztwo} spin liquids in two space dimensions; such transitions have been argued to equivalent to the condensation of an XY field coupled to a {\Uone} gauge field, where the critical XY field represents a singlet combination of spinons. A second case is the transition from a {\Uone} to a {\Ztwo} spin liquid, here in three space dimensions, which is driven by the condensation of \emph{pairs} of gauge-charged particles, akin to superconducting pairing.

To our knowledge, detailed experimental studies have not been performed for any of these transitions, mainly due to the lack of suitable materials.

\subsection{Field-driven transitions and BEC phenomena}
\label{sec:bec}

Local-moment magnets can display a variety of QPTs as function of applied magnetic field. The simplest case is the transition at the saturation field of an \SUtwo-symmetric Heisenberg magnet: Upon lowering the field, a high-field magnon becomes soft at a particular wavevector, and the transition can be understood as magnon Bose-Einstein condensation (BEC) which turns the fully polarized state into a canted antiferromagnet. The latter breaks the {\Uone} spin rotation symmetry about the field axis and is therefore also understood as a spin superfluid. The boson condensation nature of the QPT implies that this is in the universality class of the dilute Bose gas, with $z=2$ \cite{ssbook}. A similar field-driven transition occurs between the low-field singlet and intermediate-field canted phases of coupled-dimer magnets \cite{gia08}.

While these transitions involve only trivial magnetization plateaus at $M/\Msat=0$ and $1$, frustrated magnets often display intermediate magnetization plateaus. The QPTs in and out of such a magnetization plateau may be of BEC type, but are more complicated if the plateau phase spontaneously breaks lattice translation symmetry. Then, the plateau phase and the adjacent canted phase break different symmetries, possibly resulting in two continuous transitions with an intermediate coexistence (i.e. supersolid) phase or a first-order transition \cite{laflorencie07}. Experimentally, such field-induced supersolidity has been discussed for the Shastry-Sutherland compound SrCu$_2$(BO$_3$)$_2$ \cite{matsuda13} and for the spinel MnCr$_2$S$_4$ \cite{tsurkan17}.

Strong frustration often renders the magnon bandwidth small, paving the way for more exotic field-driven transitions. As has been discussed for a variety of frustrated Heisenberg models, it is possible that the high-field phase displays multi-magnon bound states whose minimal energy lies below that of the single-magnon branch. Then, upon lowering the field, the first instability is in this multi-magnon sector, and the resulting ordered state can be understood as a condensate of magnon bound states \cite{shannon06}.
The most important case is that of two-magnon bound states whose condensation induces a spin-nematic state: This is a state with quadrupolar order whose order parameter is a traceless rank-$2$ tensor. The QPT from the high-field state is either continuous of BEC type, with $z=2$, or is of first order due to large fluctuations.

Last not least, we note that spin-orbit coupling drastically modifies the physics described above. First, magnetization is no longer conserved, such that the fully polarized state is not an eigenstate of the Hamiltonian. As a result, the magnetization in the high-field phase is not saturated even as $T\to 0$. Second, the lower symmetry typically implies that field-driven transitions break discrete symmetries only. The corresponding QPT are then of $Z_n$ type, with dynamic exponent $z=1$.

\subsection{Fluctuation-induced first-order transitions}
\label{sec:brazov}

A remarkable aspect of near-critical fluctuations, well known from the physics of classical critical phenomena, is that they can render a transition discontinuous. This happens in cases where the phase space available for the critical fluctuations is exceedingly large: Then, the system may realize a first-order transition into the ordered phase which preempts the approach to criticality and hence avoids the large entropy associated with the critical fluctuations \cite{weinberg73,halperin74,bak76}. On a technical level, fluctuation effects induce either a negative quadratic coefficient or additional non-analytic terms in the Ginzburg-Landau free energy which in turn cause first-order behavior.

This physics is particularly important for frustrated magnets, because strong frustration implies large degeneracies and a weak selection mechanism for actual ordered states, both enhancing the phase space for fluctuations. One instructive theoretical scenario is that of Brazovskii \cite{brazovskii} who discussed critical fluctuations becoming soft on a finite manifold in momentum space (as opposed to a single point): This results generically in a fluctuation-driven first-order transition. Interestingly, the helimagnet MnSi, where cubic anisotropies provide a weak selection of an ordering wavevector, has been argued to realize a finite-temperature first-order transition of Brazovskii type \cite{janoschek13}.

The general considerations about fluctuation-induced first-order behavior apply to QPTs as well. Hence, one may expect that numerous QPTs in frustrated systems are driven first order. Theoretically, this has been discussed for a few models, although reliable numerical data are scarce. One example is the quantum transition between a {\Uone} spin liquid and an antiferromagnetic on the pyrochlore lattice: This has been predicted to be of first order due to fluctuations \cite{makhfudz14}, which appears consistent with QMC results obtained for a hardcore-boson model on this lattice \cite{banerjee08}. Fluctuation-induced first-order behavior has also been discussed in the context of models for deconfined quantum criticality, see Sec.~\ref{sec:dqcp}.
A quantum version of the Brazovskii theory has been discussed by Schmalian and Turlakov \cite{turlakov04} who find fluctuation-induced first-order behavior along with a quantum tricritical point.
Clearly, the abundant fluctuations near a weak first-order quantum transition may induce non-trivial thermodynamic and transport properties, but detailed studies of this physics are lacking to our knowledge.

\subsection{Deconfined quantum criticality}
\label{sec:dqcp}

An interesting scenario for unconventional transitions between symmetry-broken states is that of deconfined quantum criticality \cite{dqcp1}. It describes the possibility of a direct generic continuous QPT between two ordered states which break different symmetries -- according to Landau theory and without fine-tuning, such a transition is forbidden, as it would be either of first order or split into two continuous transitions.
At a deconfined quantum critical point, the critical degrees of freedom are fractionalized particles, and the order parameters of both phases are composites of these particles. This automatically leads to large anomalous exponents for order-parameter correlations.

\begin{figure}
\center
\includegraphics[width=0.8\linewidth]{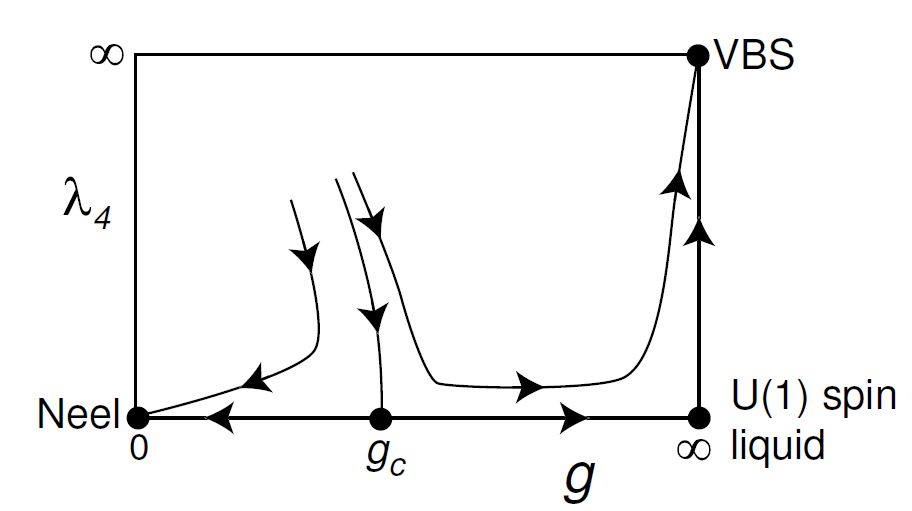}
\caption{
Schematic renormalization group flow proposed for the transition between a N\'{e}el antiferromagnet and a VBS in an {\SUtwo}-symmetric magnet, as realized e.g. by the square-lattice spin-$1/2$ $J$-$Q$ model. Increasing $g$ destabilizes magnetic order; the parameter $\lambda_4$ represents the fugacity of monopoles in the {\Uone} gauge field. The horizontal axis $\lambda_4=0$ corresponds to a non-compact CP$^1$ theory.
(Figure taken from Ref.~\cite{dqcp3})
}
\label{fig:dqcp}
\end{figure}

The most thoroughly studied instance of deconfined quantum criticality is the transition between a N\'{e}el-ordered antiferromagnet and a valence-bond solid on the square lattice. The proposed field theory employs a CP$^1$ representation of spins, with deconfined bosonic spinons and a compact {\Uone} gauge field at the critical point. The primary transition is that between a {\Uone} spin liquid and a N\'{e}el antiferromagnet, driven by the condensation of spinons which induces confinement via a Higgs mechanism; at this transition the gauge field can be assumed to be non-compact. The {\Uone} spin liquid itself is unstable towards a dimerized confined VBS phase via condensation of gauge-field monopoles, Fig.~\ref{fig:dqcp}. Hence, deconfined spinons exist only at criticality \cite{dqcp1,dqcp2,dqcp3}.

The above proposal has been tested in detailed numerical simulations of the so-called $J$-$Q$ model on the square lattice, where $Q$ denotes the strength of a ring-exchange term \cite{sandvik07}. While these simulations have verified a large part of the phenomenology of deconfined quantum criticality \cite{sandvik07,melko08}, they have also found evidence for large logarithmic corrections to scaling which are not predicted by the field-theoretical framework \cite{sandvik10}. We also note that direct numerical simulations of the proposed CP$^1$ field theory have found indications for the transition being weakly first order \cite{kuklov08}, a tendency which could not be confirmed in the $J$-$Q$ model simulations. The reasons for these discrepancies in numerical results are open, see Ref.~\cite{chalker15} for a discussion. A recent theory paper \cite{sandvik16} proposes a partial resolution by emphasizing the importance of two length scales which diverge at criticality: In addition to the spin correlation length $\xi$, there is a faster diverging scale $\xi'$ which measures the thickness of VBS domain walls \cite{dqcp2}. It appears that $\xi'$ governs the finite-scaling even of magnetic properties that are sensitive only to $\xi$ in the thermodynamic limit \cite{sandvik16}.

Recent developments in the context of field-theoretical dualities have led to additional insights \cite{nahum17}. It has been conjectured that the non-compact CP$^1$ model is dual to a so-called QED$_3$ Gross-Neveu model at criticality, the latter describing Dirac fermions coupled to both a {\Uone} gauge field and local Ising degrees of freedom. This duality suggests that the deconfined QCP between a N\'{e}el antiferromagnet and a VBS displays an emergent {\SOfive} symmetry, which is supported by numerical results \cite{chalker15}.

In addition to the N\'{e}el-VBS transition, various other Landau-forbidden transitions between two differently ordered phases have been discussed in the context of deconfined criticality. For instance, the transitions between a {\Ztwo} spin liquid and a VBS discussed in Sec.~\ref{sec:sltrans}, as well as a transition between a {\Ztwo} spin liquid and a N\'{e}el state, also belong to this class, as a {\Ztwo} spin liquid displays topological order. Emergent higher symmetries, which can be rationalized via suitable dualities, appear to be common to many of the deconfined critical points \cite{nahum17}.
To our knowledge, a clear-cut experimental example realizing deconfined quantum criticality is lacking.

\setcounter{footnote}{0}

\section{Metallic frustrated magnets}
\label{sec:frumet}

We now turn our attention to metallic solids. The presence of low-energy conduction electrons complicates the theoretical discussion, and the notion of frustration is less well defined compared to the insulating case. In this section, we summarize a number of conceptual aspects of frustrated metals, while quantum criticality in this setting will be discussed in Sec.~\ref{sec:critmet}. The discussion will mainly take a theory perspective, but connections to experiments will be highlighted when appropriate.

\subsection{Fermi liquids, non-Fermi liquids, and fractionalized Fermi liquids}
\label{sec:nfl}

A key question is whether a given metallic phase follows the Fermi-liquid phenomenology or whether it represents a genuine non-Fermi liquid -- recall that we discuss systems in $d\geq 2$ where weak electron--electron interactions do not generically produce non-Fermi liquid behavior.\footnote{At this point we ignore the potential low-temperature instability of the metal towards superconductivity. Such an instability often exists, but an analysis of the underlying normal state is indispensable for a phenomenological understanding.}

The Fermi-liquid concept requires a one-to-one correspondence of the low-energy states between the interacting system under consideration and a hypothetical system of non-interacting electrons. This implies in particular the existence of quasiparticle excitations with charge $\pm e$ and spin $1/2$ (and forbids the existence of other low-energy excitations!). It also implies the existence of a Fermi surface, defined by the momentum-space location of poles of the single-particle Green's function at energy $\w=0$. This Fermi surface then obeys Luttinger's theorem, i.e., has a momentum-space volume given by the total density of electrons $n_{\rm tot}$ (modulo filled bands):
\begin{equation}
\label{luttinger}
\mathcal{V}_{\rm FL} = K_d (n_{\rm tot}\,{\rm mod}\,2)
\end{equation}
where factors of $2$ account for spin degeneracy, i.e. a full band corresponds to $n=2$, and $K_d = (2\pi)^d / (2 V_0)$ where $V_0$ is the unit-cell volume \cite{oshi00}.
Under these conditions, the standard low-temperature Fermi-liquid properties $C(T)=\gamma T$, $\rho(T)=\rho_0+AT^2$ etc., with $\gamma$, $A$ being constants, follow immediately.\footnote{A $T^2$ behavior of the resistivity requires the existence of Umklapp scattering processes, i.e., a sufficiently large Fermi surface.}

Violation of Fermi-liquid behavior at low temperature can have various sources. In clean systems, interactions effects can produce stable non-Fermi-liquid phases. One scenario is that the low-energy excitations display quantum numbers different that of from electron or holes, leading to distinct low-temperature properties. While such behavior is generic and well understood in $d=1$, resulting in Luttinger liquids with spin-charge separation, similarly controlled descriptions in higher dimensions are scarce. A viable route to spin-charge-fractionalized metals is the doping of spin liquids, to be discussed below.

Another scenario for stable non-Fermi liquids in $d\geq 2$ has been termed fractionalized Fermi liquid \cite{flst1,flst2}. In such a phase, charged excitations have conventional quantum numbers (charge $\pm e$ and spin $1/2$), but these coexist with additional fractionalized degrees of freedom. A generic construction starts from a fractionalized spin liquid and adds conventional carriers in a second band. If these subsystems remain weakly coupled, they realize a FL$^\ast$ phase. Importantly, such a phase displays a Fermi surface with a volume violating Luttinger's theorem \eqref{luttinger} in a quantized fashion, often \cite{flst1}
\begin{equation}
\label{lutt_flst}
\mathcal{V}_{{\rm FL}^\ast} = K_d ((n_{\rm tot}-1)\,{\rm mod}\,2)
\end{equation}
where the $-1$ accounts for the electrons forming the spin-liquid component. Low-temperature properties may or may not be Fermi-liquid-like, depending on whether the emergent excitations of the spin-liquid component are gapped or gapless. Fractionalized Fermi liquids may display a variety of instabilities driven by the strong correlations in the local-moment sector, including unconventional superconductivity \cite{flst1,urban}.

\subsection{Conventional order vs. topological states}

Metals may display symmetry-broken states in a qualitatively similar manner as local-moment insulators. A conceptual difference concerns the symmetric case: Whereas a fully symmetric state of a local-moment system with half-odd-integer spin per unit cell is generically fractionalized -- see the discussion on the Lieb-Schultz-Mattis-Hastings theorem in Sec.~\ref{sec:qsl} -- realizing a fully symmetric metal is much simpler: this is just the familiar Fermi liquid.

There are less trivial symmetric states, with fractionalized Fermi liquids as well as other doped spin liquids belonging to this class. Given the insights into topological properties of fractionalized insulating phases, one may wonder about the topological characterization of non-Fermi-liquid metals. To our knowledge, relatively little work has been done in this direction. A sharp distinction between FL and FL$^\ast$ is the Fermi volume, and this can be considered a topological distinction. In contrast, some of the indicators established for insulators, like ground-state degeneracies and entanglement, cannot be easily applied because of the absence of an excitation gap \cite{urban}, and more work is needed to clarify the topological nature of non-Fermi liquid metals.

\subsection{Frustrated Hubbard models at intermediate coupling}

Given the rather detailed knowledge of the behavior of local moments on geometrically frustrated lattices, it is natural to consider metallic states on such lattices. As local-moment systems are Mott systems driven insulating by a large on-site Coulomb repulsion, a reduction of this Coulomb repulsion generically causes a Mott insulator-to-metal transition in the case of a half-filled band. Close to the Mott transition, the metallic state will be strongly correlated: It will display Hubbard bands in the single-particle spectrum which signal local-moment formation at intermediate scales, and these local moments will be subject to frustration.

A fully systematic understanding of such frustrated metals is lacking to date. Assuming that the low-temperature state is a Fermi liquid, it is commonly assumed that frustration will further (in addition to the Coulomb interaction) reduce the bandwidth of coherent quasiparticle dynamics and correspondingly lead to a much reduced coherence scale, resulting in heavy-fermion-like behavior. This reasoning is supported by numerical results, obtained using cluster generalizations of dynamical-mean-field theory (DMFT), variational cluster methods, and path-integral renormalization group methods, e.g., for the triangular \cite{imada02,tremblay06,kawakami08,gros13} and kagome lattices \cite{kawakami06} as well as a frustrated cubic lattice \cite{laubach16}.

\subsection{Doped frustrated Mott insulators}
\label{sec:dopedmott}

An alternative way to turn a frustrated Mott insulator into a metal is by doping, i.e., by varying the band filling away from half-filling. Here we remind the reader that the physics of doped Mott insulators is not well understood even without frustration, and constitutes an active and challenging field of research, with the prime experimental application being cuprate high-temperature superconductors. Geometric frustration adds to the complexity of the problem, and only a few results are established beyond doubt.

It is conceivable that large doping levels away from half-filling, i.e., small densities of electrons or holes in an otherwise empty or full band, lead to a Fermi-liquid state because scattering events between particles become rare such that interaction effects are weak. In contrast, for small doping levels, local-moment formation and hence frustration will become important, and it has been speculated that metallic doped Mott-insulator phases with non-Fermi liquid character emerge \cite{pwa87,leermp,qiss10}. Such a non-Fermi metal can either be a fully fractionalized state with spinon and holon excitations, or it can be a fractionalized Fermi liquid with conventional charge-$e$ spin-$1/2$ quasiparticles, see Sec.~\ref{sec:nfl}.

Interestingly, a recent numerical study \cite{hcjiang17} of a kagome-lattice $t$-$J$ model indicated that, at least a small doping, the doped spin liquid does \emph{not} become metallic, as the holes form a Wigner crystal driven by the spin-singlet background. This appears qualitatively consistent with experiments on electron-doped Herbertsmithite, ZnLi$_x$Cu$_3$(OH)$_6$Cl$_2$, which remains insulating up to $x=1.8$ \cite{kelly16}. However, dopant-induced disorder which is not part of the $t$-$J$ model may add to carrier localization.

\subsection{RKKY frustration and Kondo lattices}

Multiband systems offer more possibilities to generate frustration. A frequent situation is the presence of one band of strongly correlated electrons (often 4f or 5f) which generate local moments. In the presence of other weakly correlated metallic bands, these moments are coupled via indirect Ruderman-Kittel-Kasuya-Yosida (RKKY) interaction. This interaction is determined by the low-energy bandstructure and long-ranged.
Clearly, depending on lattice geometry and bandstructure details, the RKKY interaction can be frustrated. In particular, a geometrically frustrated lattice is not a precondition for frustration, because further-neighbor interactions can counter-act magnetic order also on bipartite lattices.

The situation of RKKY-coupled local moments is generically realized in heavy-fermion materials, theoretically described by Kondo-lattice models. Here, the RKKY interaction competes with Kondo screening which by itself suppresses magnetic order \cite{doniach77}. As a result, complex cases of quantum criticality may emerge in such materials, to be discussed in Sec.~\ref{sec:hfpd} below.


\section{Quantum criticality in frustrated metals}
\label{sec:critmet}

This section is devoted to quantum criticality in metallic systems with frustration, as described in Sec.~\ref{sec:frumet}. In analogy to Sec.~\ref{sec:critins} we will distinguish different types of transitions depending on the phases involved. Two main differences with respect to insulators will arise: (i) The presence of low-energy fermionic excitations changes the nature of the critical points. (ii) Entirely novel transitions are possible which involve the (partial) onset or loss of metallicity due to interactions, i.e., Mott-type transitions.

\subsection{Conventional ordering transitions}

As with insulators, QPTs in metals may involve the onset of conventional, i.e., symmetry-breaking, order. Entering the ordered state reconstructs the Fermi surface, but leaves metallicity intact.\footnote{Two exceptions are noteworthy where the onset of conventional order coincides with a metal-to-insulator transition, namely the case of perfect nesting where the onset of order removes the entire Fermi surface, and the case of a semimetal (like graphene) where the onset of order removes the Fermi points.}

As before, the simplest theoretical description is given by the LGW approach which considers a theory for the local order parameter only. Accounting for the presence of low-energy particle hole pairs introduces Landau damping into the field theory Eq.~\eqref{phi4qu}: Integrating out the fermions perturbatively yields, to leading order, a term of the form $|\w| \op(\w)^2$ or $(|\w|/|\vec{q}|) \op(\w)^2$ and hence a dynamical exponent for the bosonic order-parameter fluctuations of $z_B=2$ or $z_B=3$ for the cases of ordering wavevector $\vec{Q}\neq 0$ (e.g. antiferromagnet) or $\vec{Q}=0$ (e.g. ferromagnet), respectively. This approach has been pioneered by Hertz \cite{Hertz76}, Millis \cite{Millis93}, and Moriya \cite{Moriya85}. Its main results are reviewed and discussed vis-a-vis experimental data in Ref.~\cite{hvl}. Importantly, the above values of $z_B$ imply that the LGW theory for $d\geq2$ is never below its upper critical dimension.

Subsequent theoretical work has shown that the LGW approach to metallic quantum criticality is in many cases not warranted, because higher-order terms introduce singularities and non-localities into an order-parameter-only field theory -- this applies in particular to ferromagnetic transitions and to all transitions in $d=2$. In principle, a consistent and tractable asymptotic theory needs to include both order-parameter fluctuations and low-energy fermions. The analysis of such theories is notoriously difficult, but partial progress has been made over the last decade.

In the ferromagnetic case with $\vec{Q}=0$, it has been argued that singular terms in the effective action, rooted in singular corrections to the spin susceptibility of a Fermi liquid, render the QPT generically first-order in both $d=2$ and $d=3$ \cite{bkv97}. Experimentally, most ferromagnetic QPTs are indeed found to be first order, exceptions being compounds with sizeable quenched disorder \cite{brando_rev}.

Other cases with $\vec{Q}=0$ include nematic order that breaks discrete lattice rotational symmetry and theories of fermions minimally coupled to a {\Uone} gauge field -- here a continuous QCP survives. An analysis in $d=3$ gives to $z_B=3$ and a self-energy $\Sigma(\w,\vec{k})\propto|\w|$, sometimes dubbed marginal Fermi-liquid behavior \cite{ssbook}. In $d=2$ the situation is more complicated: Solving the coupled problem of bosons and fermions self-consistently in a particular large-$N$ limit, with $N$ the number of fermion flavors, yields $z_B=3$ and a self-energy $\Sigma(\w,\vec{k})\propto|\w|^{2/3}$, implying the destruction of quasiparticles \cite{lee89,metlitski10a}. However, $1/N$ corrections are singular \cite{lee09}, and the full answer is not known.

For non-zero ordering wavevector $\vec{Q}$, the Fermi surface develops hot spots: These are points in momentum space which are connected by $\vec{Q}$ and at which potentially singular scattering of fermions occurs. For $d=3$ the results obtained from the LGW approach are believed to be correct, i.e., bosons and fermions remain weakly coupled at the QCP. In $d=2$ a self-consistent large-$N$ theory  yields $z_B=2$, with a hot-spot fermionic self-energy of $\Sigma(\w,\vec{k}_{HS})\propto|\w|^{1/2}$ \cite{abanov_rev03}. However, this theory displays a flow to smaller $z_B$ at finite $N$ again from singular $1/N$ corrections \cite{metlitski10b}. A very recent work \cite{schlief17} suggested that the correct asymptotic result is $z_B=1$ instead, accompanied by a self-energy $\propto |\w|$. On the numerical front, recent Monte-Carlo results \cite{schattner17} for antiferromagnetic metallic quantum criticality appear, however, more consistent with $z_B=2$. This disagreement is not understood, but could be related to a slow flow from $z_B=2$ to $z_B=1$.

We finally note that theory work in the context of heavy-fermion metals has proposed an alternative scenario for the $d=3$ antiferromagnetic transition \cite{abrahams14} which involves strong coupling between bosonic order-parameter fluctuations and fermions. In this semi-phenomenological approach, non-Fermi liquid behavior emerges as a result of energy fluctuations at small wavevector, leading to critical quasiparticles on the entire Fermi surface, while the non-linearities in the bosonic sector remain perturbative.
Given that the weak-coupling LGW theory appears to be internally consistent as well, one would then expect the existence of a multicritical point controlling the transition between weak-coupling and strong-coupling antiferromagnetic quantum criticality, but a theory for this is not known.
Remarkably, the results of the strong-coupling theory, such as hyperscaling with $z_B=4$, $\nu=1/3$ in $d=3$ \cite{abrahams14}, appear to match experimental data obtained for \yrs; however, alternative scenarios for this material have been proposed as well \cite{steglich10}.
In summary, metallic quantum criticality in general constitutes an important problem which is only partially understood.

In frustrated systems, additional complications arise, partially discussed already in Section~\ref{sec:critins} for insulators. These include higher (emergent) order-parameter symmetries at criticality, soft ordering wavevectors, and dimensional reduction. Moreover, the electronic bands may be anomalously flat due to frustration, generating additional small energy scales. Then, reaching the asymptotic critical regime requires lower temperatures, and the intermediate-temperature physics is dominated by broad crossover regimes -- these are discussed in Sec.~\ref{sec:broadxover}.

\subsection{Mott transitions and partial Mott transitions}

\begin{figure}
\center
\includegraphics[width=0.8\linewidth]{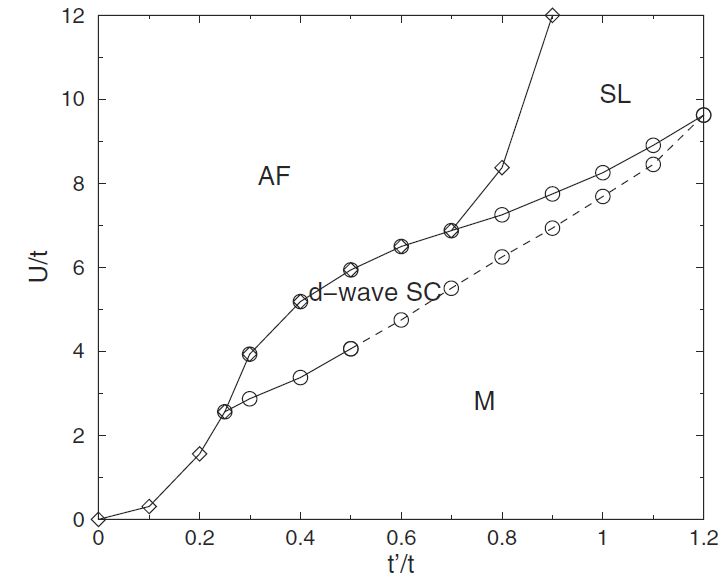}
\caption{
Cluster-DMFT phase diagram of the Hubbard model on an anisotropic triangular lattice as function of Hubbard interaction $U$ and hopping ratio $t'/t$, where $t'=0$ and $t'=t$ correspond to the square and triangular lattices, respectively. M, SC, AF, SL denote metal, superconductor, antiferromagnetic insulator, and spin-liquid phases, respectively. Solid (dashed) lines correspond to first-order (continuous) transitions.
(Figure taken from Ref.~\cite{tremblay06})
}
\label{fig:triang}
\end{figure}

A Mott transition is an interaction-driven metal-to-insulator transition: It transforms a half-filled metallic band into an insulator of local moments. The Mott-insulating state is often accompanied by antiferromagnetic long-range order, and the quantum transition from a paramagnetic metal to an antiferromagnetic Mott insulator is generically of first order (or involves an intermediate antiferromagnetic metallic phase).
This is different in the case of a spin-liquid Mott insulator: A ``genuine'' zero-temperature Mott transition from a paramagnetic metal to an insulating spin liquid can be continuous. As the existence of the spin liquid requires frustration, such transitions are expected to occur in half-filled Hubbard models on frustrated lattices upon varying $U/t$. In fact, a metal-to-spin liquid transition has been found in numerical simulations of the triangular-lattice Hubbard model which, however, appears to be first order \cite{imada02,tremblay06,kawakami08}, Fig.~\ref{fig:triang}, with superconductivity possibly appearing on the metallic side before the Mott transition \cite{tremblay06}. An candidate experimental realization is in the organic compound {\kaETCN} under pressure \cite{shimizu03}.

A defining criterion for a Mott transition is a quantized change in the Fermi volume: In a Fermi liquid, the momentum-space volume enclosed by the Fermi surface is given by the total number of electrons according to Luttinger's theorem \eqref{luttinger}. In a Mott insulator, there is no Fermi surface\footnote{We do not consider the so-called Luttinger volume, $V_{\rm lutt} = \int_{G(k)>0} dk$, which accounts for both poles and zeroes of the Green's function. For an in-depth discussion on aspects of the Luttinger volume in Mott insulators see Ref.~\cite{rosch07}.}, and hence the Fermi volume changes at a single-band Mott transition by $K_d \times 1$. Such an abrupt change is nevertheless compatible with the QPT being continuous: Upon approaching a continuous Mott transition from the metallic side, the quasiparticle weight on the Fermi surface will vanish continuously, while the charge gap opens continuously on the insulating side. At criticality, one expects a \emph{critical Fermi surface}, i.e. a well-defined $(d\!-\!1)$-dimensional manifold in momentum space where the electronic spectral function displays (possibly momentum dependent) power-law singularities \cite{senthil_critfs}.

In a multi-band system, there is the possibility for a partial Mott transition. This is a transition between two {\em metallic} phases where the Fermi surface undergoes a quantized change. In the simplest case, one band (or orbital) changes its character from metallic to Mott-insulating while other bands remain metallic. Consequently, such a transition has also been dubbed orbital-selective Mott transition \cite{anisimov02,pepin07,osmottrev}. If stable in the low-temperature limit, the partial Mott phase violates Luttinger's theorem \eqref{luttinger} and, hence, is a non-Fermi liquid metal. This is precisely the fractionalized Fermi-liquid phase (FL$^\ast$) introduced in Sec.~\ref{sec:nfl} above, and a transition between FL and FL$^\ast$ is an orbital-selective Mott transition. Phenomenologically, such a transition can be expected to be accompanied by a jump in the Hall constant \cite{coleman05}.

A concise theoretical understanding of continuous zero-temperature Mott (or partial) Mott transitions is lacking to date. Most theoretical descriptions are based on slave-particle theories which involve separate degrees of freedom representing spin and charge of the electrons. Often, the charge degrees of freedom are encoded by bosons which are gapless and condensed in the metal, but gapped and disordered in the insulator. Hence, the insulator-to-metal transition becomes a BEC transition of charged bosons coupled to a gauge field \cite{flst2,senthil_mott}. However, such a description (at least in its simplest version) does not account for possible non-trivial momentum dependencies along the Fermi surface. Moreover, the fermionic character of the Mott phenomenon might require a formulation using non-bosonic critical degrees of freedom, but to our knowledge a successful theory of this type has not been formulated.

Two further aspects are worth mentioning:
(i) A partial Mott transition has been deduced from numerical results for the doped single-band Hubbard model in the context of cuprate superconductors. Here, the insulating behavior appears to be momentum-selective, i.e., some regions in momentum space behave Fermi-liquid-like while others behave Mott-like. This transition has been interpreted in terms of a doping-driven FL$^\ast$--FL transition, and parallels between heavy-fermion materials and cuprates have been discussed \cite{ss12}. A possible effective theory, involving the condensation of a Higgs field corresponding to fluctuating antiferromagnetic order, has been proposed in Ref.~\cite{chowdhury15}, but has not been established beyond doubt.
(ii) Apparent quantum critical behavior at elevated temperatures has been detected above the finite-temperature endpoint of a first-order Mott transition line. This remarkable observation, manifest, e.g., in scaling behavior of the resistivity, was first made in DMFT simulations of the single-band Hubbard model on a Bethe lattice \cite{terletska11}, and later verified experimentally in three pressure-tuned organic compounds \cite{kanoda15}. It is tempting to speculate that this behavior arises from a proximate quantum critical point of genuine Mott type, but a deeper theoretical understanding is lacking at present.

\subsection{Global phase diagram of heavy fermions}
\label{sec:hfpd}

Heavy-fermion metals form a particular fertile ground for metallic quantum criticality, because they host a multitude of ordered phases and are easily tunable \cite{hvl,steglich10}. As pointed out early on by Doniach \cite{doniach77}, the heavy-fermion phase diagram is governed by the competition between the Kondo coupling and RKKY interactions between local moments, leading to heavy Fermi liquids and ordered magnetic states, respectively. Later on, it has been suggested \cite{flst2,si_global,mv_af,si10,coleman_global} to consider, in addition to the ratio between Kondo temperature and RKKY interaction, a second tuning parameter which acts to suppress magnetic order in the local-moment subsystem -- this is loosely labelled as ``frustration'' (alternatively: ``quantum fluctuations''). This tuning parameter naturally enables access to fractionalized states.

\begin{figure}
\center
\includegraphics[width=0.8\linewidth]{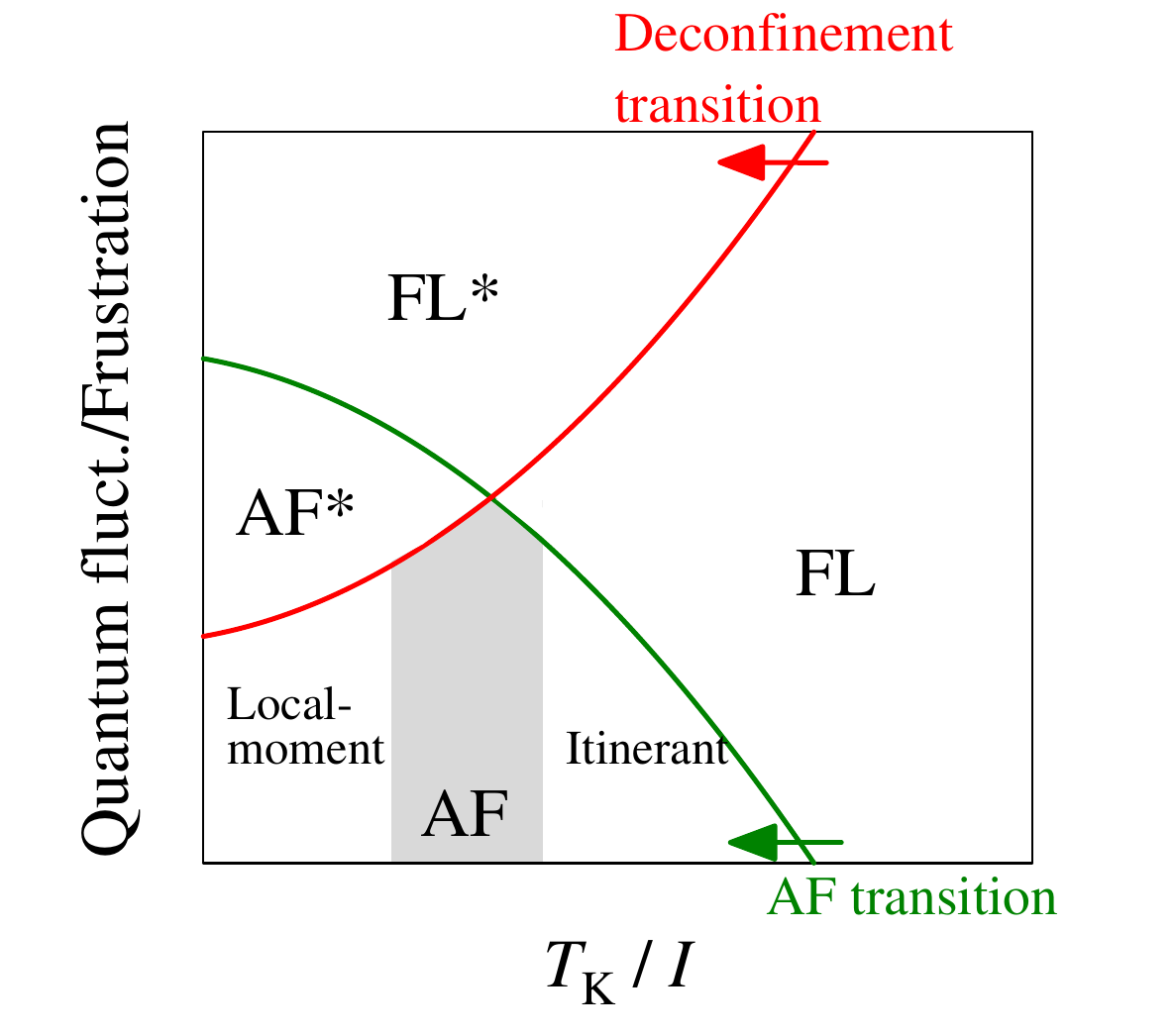}
\caption{
``Global'' phase diagram for heavy-fermion metals (with one $f$ electron per crystallographic unit cell), with two transitions for the onset of antiferromagnetism and for the breakdown of the Kondo effect (equivalently the onset of deconfinement). FL$^\ast$ is the fractionalized Fermi-liquid phase described in Sec.~\ref{sec:nfl}. Inside the AF phase, a crossover from more itinerant to more localized behavior occurs, which may be accompanied by one or more transitions where the Fermi-surface topology
changes. Lastly, AF$^\ast$ refers to a fractionalized magnet, with magnetic LRO and fractionalized excitations coexisting.
(Figure taken from Ref.~\cite{mv_af})
}
\label{fig:global}
\end{figure}

The resulting ``global'' phase diagram of heavy fermions is shown in Fig.~\ref{fig:global}. It features two transition lines, one involving the onset of antiferromagnetism and one involving the onset of deconfinement. Importantly, the onset of deconfinement in the paramagnetic metallic phase corresponds to an orbital-selective Mott transition into an FL$^\ast$ phase as advocated above, as FL$^\ast$ features deconfined fractionalized excitations in the local-moment sector. Such an orbital-selective Mott transition is easily driven by the reduction of Kondo screening in a frustrated regime, because it is Kondo screening which renders the local-moment electrons metallic. Hence, the onset of deconfinement also corresponds to a breakdown of the Kondo effect.
The two transition lines define four phases: In addition to the paramagnetic phases FL and FL$^\ast$, there are a conventional (AF) and a fractionalized (AF$^\ast$) antiferromagnet. The Fermi volume is ``large'' in the FL phase, i.e., encloses both conduction and local-moment electrons, while it is ``small'' in FL$^\ast$ because it is determined by conduction electrons alone, hence violating Luttinger's theorem. In the metallic AF and AF$^\ast$ phase, translation symmetry breaking enlarges the unit cell, such that Luttinger's theorem is generically fulfilled. The transition from FL to AF is hence a conventional ordering transition, accompanied by the backfolding of bands.

A slightly different version of the global phase diagram has been put forward in Ref.~\cite{si_global,si10}, the main difference being that the coincidence of the Kondo-breakdown and magnetic transition lines is not considered accidental, but systematic. Ref.~\cite{si0103} has developed a corresponding extended DMFT description of a Kondo breakdown driven by magnetic criticality. Alternatively, this might be viewed as a case of deconfined criticality \cite{flst3}.

Experimentally, magnetic quantum criticality has been investigated in numerous heavy-fermion materials, and we refer the reader to Refs.~\cite{stewart01,hvl,gegenwart16} for an extensive review. In a number of cases, the observed critical behavior is consistent with the predictions from the LGW approach for antiferromagnetic transitions in metals in $d=3$, and hence corresponds to the FL--AF transition in Fig.~\ref{fig:global}. However, in other cases the critical singularities appear stronger, most notably in {\ceau}, {\yrs}, and {\ybal}. This has been suggested to be related to Kondo-breakdown physics, but strong-coupling effects, multi-criticality or further, yet unexplored effects may also play a role.

We finally note that the metallic phases discussed here can in principle be unstable towards superconductivity. Pairing tendencies appear to be particularly strong near the quantum phase transitions: Critical magnetic fluctuations have been shown to be source of strong cooper pairing on general grounds \cite{abanov_rev03,chubukov13}, with results for Kondo lattice models \cite{becca14} being consistent with this idea. In addition, the transition between a {\Ztwo} FL$^\ast$ and FL phases has been argued to display a generic pairing instability due to the spinon pairing inherent to a {\Ztwo} spin liquid \cite{flst1}.
Experimentally, a number of heavy-fermion compounds display a superconducting dome near their antiferromagnetic QCP, with CeIn$_3$ and CePd$_2$Si$_2$ being prominent examples \cite{mathur98}.


\section{Further ingredients}
\label{sec:ingredients}

\subsection{Broad crossover regimes}
\label{sec:broadxover}

In addition to the possibly exotic physics in the low-temperature limit discussed so far, an issue of eminent experimental relevance is the existence of very broad crossover regimes in frustrated systems. As noted in Sec.~\ref{sec:frumet}, frustration has the general tendency to reduce energy scales -- this is manifest in electron or magnon bands of reduced width and reduced coherence scales.

Many continuum-limit field theories describing quantum critical behavior only apply at energies and temperature {\em below} these emergent low-energy scales, such that reaching the asymptotic low-temperature regime may be very difficult experimentally.\footnote{Even without the complication of frustration, observables may display large subleading corrections to their leading scaling behavior, such that measurements over a wide range of energies or temperatures, far below all microscopic scales, are required to extract leading power laws. This applies to both experiments and computer simulations.} As a result, measurements may reflect crossover instead of critical behavior. For frustrated systems, such crossover behavior can be very interesting as well, e.g., governed by degenerate manifolds -- this is, however, usually not quantum critical behavior in the sense of scale invariance in space and time.

The issue of possibly broad crossover regimes induced by small microscopic energy scales is, in fact, a known complication for heavy-fermion metals where the effective Fermi energy is set by the Kondo temperature, typically of order $10$--$100$\,K. If, in addition, the bands display fine structure in momentum space, then the relevant low-energy scales dictated by bandstructure can be easily be as low as a few Kelvin, and asymptotic critical behavior can only be expected significantly below these scales.

\subsection{Quenched disorder and glassiness}

Quenched disorder, being inevitably present in every real solid in form of structural or substitutional defects, can strongly influence the low-energy behavior of a given system. This applies in particular to frustrated systems because of their tendency to phase competition and large low-temperature entropies.

Quite generically, the combined effect of disorder and frustration tends to induce ``glassy'' behavior \cite{villain79}. Originally, this term refers to the presence of a broad spectrum of relaxation times, leading to extremely slow dynamics. Such glassy dynamics is caused by an energy landscape with a huge number of inequivalent and almost degenerate minima. In canonical spin glasses, these minima represent states without conventional long-range order, i.e., correlation functions decay exponentially. Nevertheless, it is believed that one can define a thermodynamic spin-glass phase, where the static expectation values of individual spins are non-vanishing, signalled by a finite Edwards-Anderson order parameter \cite{sg_book}.

In practice, frustrated spin systems with quenched disorder often feature a short-ranged glassy version of the order of the corresponding clean system.\footnote{Some highly frustrated system with a classically degenerate ground-state manifold display a phenomenon dubbed ``order by quenched disorder'' where the presence of defects tends to select a particular state from the degenerate manifold which may be \emph{different} from that selected by thermal or quantum fluctuations. One example is the triangular-lattice Heisenberg model in a magnetic field \cite{maryasin13}.} Spins freeze, i.e., cease to fluctuate, below a freezing temperature $\Tf$. Glassiness is signalled by broadened magnetic Bragg peaks and by a response which depends on both history and frequency of an applied perturbation, i.e., a splitting between field-cooled and zero-field-cooled susceptibilities and a frequency-dependent AC susceptibility below $\Tf$. In addition, the specific heat does not display a sharp anomaly at $\Tf$ but only a weak and broad maximum above $\Tf$, as is typical for spin glasses. Sometimes such states are called ``cluster spin glasses'', due to the finite (often sizeable) correlation length.

Consider now a QPT involving such a disorder-induced glassy state. Clearly, the quantum-critical behavior at short lengths and elevated energies will be that of the clean system, but the asymptotic low-energy long-wavelength physics will be dominated by glassiness. Unfortunately, very little is known about quantum spin glasses and their criticality, with the exception of a few toy models \cite{oppermann95,parcollet00}. Consequently, this constitutes an important (but notoriously difficult) avenue of future research.

We note that there exists an established and growing body of work which deals with the influence of quenched disorder on continuous phase transitions -- both classical and quantum -- involving the onset of LRO, and we refer the reader to Refs.~\cite{TV} for reviews. One element of this theory is a criterion due to Harris \cite{Harris74,CCFS86} which states that a given transition is stable under the influence of small disorder provided that its correlation-length exponent obeys $\nu > 2/d$. If this condition is violated, the universality of the transition changes due to disorder, or the transition disappears entirely. A more thorough classification for QPTs under the influence of disorder has been developed recently \cite{TV}.
However, most of these theoretical considerations assume that the phases on both sides of the transition are itself stable against quenched disorder. It is precisely this condition which is often violated for frustrated systems because of disorder-induced glassiness; in such a situation the Harris criterion and its consequences are not of direct relevance.

\subsection{Quantum multi-criticality}

Given that frustration acts to suppress simple forms of order, it is clear that it also promotes the competition of {\em multiple} alternative states: A frustrated spin system may feature instabilities to both valence-bond solid and unconventional magnetic order, and a frustrated metal may display tendencies to both nematic order and weak antiferromagnetism.

The presence of multiple low-temperature instabilities paves the way to consider quantum multi-criticality. While multi-critical points in general require double fine-tuning to be reached, it is conceivable that a material displays two nearby quantum critical points, associated e.g. with different ordering phenomena, such that experiments at elevated temperature detect this as a single multicritical point. Note that this scenario is very different from deconfined quantum criticality, described in Sec.~\ref{sec:dqcp}, which also involves two different ordering phenomena emerging from a single critical point, but is characterized by fractionalized critical degrees of freedom.

Theoretically, quantum multi-criticality in the framework of Hertz-Millis theory has been recently studied in Ref.~\cite{gregory16}. Experimentally, the metals NbFe$_2$ \cite{nbfe2} and {\yrs} \cite{hvl,yrs} have been discussed as candidate materials, because both display unusual quantum critical behavior, agreeing with the predictions for critical antiferromagnetism for some observables and with those for critical ferromagnetism for others.

\subsection{Orbital degrees of freedom and frustration}

So far, we have employed the notion of frustration or magnetic degrees of freedom. Correlated electron systems often also involve orbital and/or charge and/or lattice degrees of freedom which, in principle, can be frustrated as well. Orbital frustration can lead to interesting new phenomena, which we quickly discuss in the following \cite{feiner97,mostovoy03,khaliullin05}.

Orbital degrees of freedom arise for lattice structures where the crystalline electric field leads to degenerate orbitals which are partially filled \cite{tokura00}. For instance, in a cubic environment the $t_{2g}$ states of the $d$ shell are degenerate, such that e.g. a $d^1$ ion carries, in addition to a spin $1/2$, also a threefold orbital degeneracy. For Mott insulators, the combined physics of spins and orbitals is often described in Kugel-Khomskii-type models \cite{kugel82}, written down in terms of spins and orbital pseudospins with near-neighbor interactions. We note, however, that the underlying symmetry of the orbital part is often lower than that of the spin part.

Depending on the lattice structure, it is clear that frustration may also arise in the orbital sector, and often intertwines in a non-trivial way with that of the spin sector. Different interesting scenarios have been discussed \cite{mostovoy03}, for instance:
(i) Orbital order can relieve magnetic frustration, by strengthening or weakening certain exchange paths. This typically results in states which display both long-range orbital and spin order, with NaVO$_2$ being a nice example \cite{mcqueen08}.
(ii) Spins and orbital pseudospins can conspire to produce a spin-orbital liquid. In fact, an orbital liquid state was proposed in early work for LaTiO$_3$ \cite{khaliullin00}; however, this material turned out to display a Jahn-Teller distortion lifting the orbital degeneracy \cite{cwik03}. More recently, the A-site spinel FeSc$_2$S$_4$ has been suggested to realize an entangled spin-orbital singlet state, proximate to a QCP to an orbital-ordered antiferromagnet \cite{balents08}. Recent experiments, however, have detected a structural distortion followed by weak magnetic order at low temperature in this compound, placing it on the opposite side of this QCP \cite{broholm16}. The precise nature of the crossovers and the role of quenched disorder in FeSc$_2$S$_4$ are not fully understood.

Given the complexity of spin-orbital models, a significant body of theory work has also been devoted to orbital-only models, often dubbed compass models; some of these models can also be viewed as effective models for frustrated magnets with strong spin-orbit coupling. Depending on the symmetries of the involved orbitals and the underlying lattice, the behavior of these models can be very rich, and  we refer the reader to the literature \cite{compass_rmp,oles_compass} for further details.

\subsection{Charge frustration}

In lattice systems with mobile charges, the concept of frustration can as well apply to charge degrees of freedom. If the latter live on a frustrated lattice and are strongly interacting, states akin to spin liquids may form.
This is most easily seen when considering spinless fermions where the charge state on a site can be interpreted as an Ising pseudospin. Nearest-neighbor repulsion of charges, $V$, corresponds to an antiferromagnetic Ising interaction, and charge hopping $t$ implements quantum fluctuations. Importantly, fixed particle number now implies fixed magnetization in the spin language, i.e., the states are more conveniently discussed at fixed magnetization as opposed to fixed field.

A number of theory ideas for charge frustration, occurring in the limit $V\gg t$, have been discussed in the literature. An interesting case is so-called charge ice, describing the physics of charges $\pm q$ on a pyrochlore lattice subject to Coulomb-type interactions. This model displays a liquid-like regime, akin to spin ice, with fractionally charged excitations $\pm q/2$ \cite{pollmann14}. Fractionalized charge excitations have also been discussed for spinless fermions on a checkerboard lattice \cite{pollmann06} as well as for spinful fermions on a partically filled pyrochlore lattice, the latter existing in a quantum spin liquid state \cite{chenkim14}.

Experimentally, charge degrees of freedom correspond to mixed-valent ions, and the continuous character of the valence implies that charge order is more flexible than Ising spin order. One interesting example of charge frustration is the spinel AlV$_2$O$_4$, with nominal valence V$^{2.5+}$, which settles into a ``three-one'' charge-ordered state below $700$\,K, with three V$^{2+}$ and one V$^{4+}$ ion per unit cell, accompanied by a rhombohedral lattice distortion \cite{matsuno01}.


\section{Experimental examples}
\label{sec:exp}

The purpose of this final section is to connect the theoretical concepts and ideas outlined in the previous sections to concrete experimental results. We will restrict ourselves to solid-state materials, not covering developments in the field of cold atomic gases and elsewhere. Even with this restriction, the number of experiments and compounds is large, and we select a small subset of them which we find particularly important or interesting -- such a choice is necessarily biased and incomplete.

\subsection{Dimensional reduction: \bacusio, \ceau}
\label{sec:bacusio}

\paragraph{\bacusio.}
The insulating magnet {\bacusio}, also known as Han purple, is a prime example of a spin-$1/2$ coupled-dimer system. It consists of square-lattice bilayers which display an AB stacking along the c-axis direction, such that dimers form a body-centered tetragonal structure \cite{bacusio}. As a result, an antiferromagnetic interaction between bilayers would be fully frustrated. In zero field, the material is a singlet quantum paramagnet, as expected for strong intradimer coupling.

Experiments have observed field-driven Bose-Einstein condensation of magnons \cite{gia08}, with the remarkable feature that the exponents of the QPT reflect 2D (instead of 3D) behavior down to lowest temperatures \cite{sebastian06}. This dimensional reduction has been interpreted in terms of interlayer frustration: In this scenario, geometric frustration of the inter-bilayer coupling renders the crossover scale from 2D to 3D behavior unobservably small. As detailed in Ref.~\cite{roesch07b}, this frustration effect may cooperate with an inter-bilayer modulation: Due to a structural phase transition at low $T$, adjacent bilayers in {\bacusio} are no longer equivalent, leading to an inhomogeneous condensate which reduces the effective c-axis coupling.

Notably, results from recent NMR experiments, combined with a re-analysis of neutron-scattering data, suggest that the inter-bilayer coupling is in fact {\em ferromagnetic} \cite{stern14}, which would rule out frustration as a source of dimensional reduction. A full theory for the behavior of {\bacusio} is thus lacking.

\paragraph{\ceau.}
CeCu$_6$ doped with Au or Ag has been a posterchild for magnetic heavy-fermion quantum criticality. {\ceau} is magnetically ordered for $x>0.1$ and displays critical singularities at $x=0.1$ such as a logarithmically divergent specific-heat coefficient \cite{hvl}. This thermodynamic critical behavior was originally interpreted in terms of LGW criticality with 2D spin fluctuations \cite{rosch97}, also fueled by neutron scattering measurements which identified critical spin fluctuations along lines in momentum space for $x=0.1$ \cite{stockert98}. The origin of this dimensional reduction is not understood.

However, various additional observations appear inconsistent with predictions from LGW theory. These include the unconventional scaling exponent deduced from $\w/T$ scaling of susceptibility data \cite{schroeder00} and the weak logarithmic divergence of the Gr\"uneisen parameter in the related compound {\ceag} at its critical point $x=0.2$ \cite{kuechler04}. A recent detailed study of the anisotropic thermal expansion of CeCu$_{5.9}$Au$_{0.1}$ \cite{grube17} has confirmed approximately logarithmic behavior in thermal expansion and Gr\"uneisen parameters, but with a pronounced anisotropy which has been related to the anisotropy of the critical magnetic fluctuations..

While a number of theory proposals have been made to explain the quantum critical behavior of {\ceau}, including Kondo-breakdown and strong-coupling magnetic criticality, it is fair to say that a complete understanding has not been reached \cite{hvl,abrahams14,steglich10}.

\subsection{Interchain frustration: CoNb$_2$O$_6$}

The mineral columbite, CoNb$_2$O$_6$, is a magnetic insulator consisting of weakly coupled Ising chains with an exchange scale of $20$\,K. Under application of a magnetic field, it represents, to leading order, a beautiful realization of the one-dimensional transverse-field Ising model, as evidenced by neutron scattering \cite{coldea10} and NMR \cite{imai14} experiments.

This 1D behavior is cut-off at low energies and temperatures by inter-chain coupling which leads to antiferromagnetic order below $2.9$\,K in zero field. Interestingly, the inter-chain coupling is frustrated because the chains are arranged in a triangular fashion, giving rise to physics akin to the triangular-lattice Ising model. Including the relevant deviations from perfect triangular geometry leads to a rich low-temperature phase diagram in an applied field, with a non-trivial interplay between frustration and criticality, theoretically studied in Ref.~\cite{kaul10}. To our knowledge, this has not been probed in detail experimentally.

\subsection{Kitaev interactions: \rucl}

\begin{figure}
\center
\includegraphics[width=0.8\linewidth]{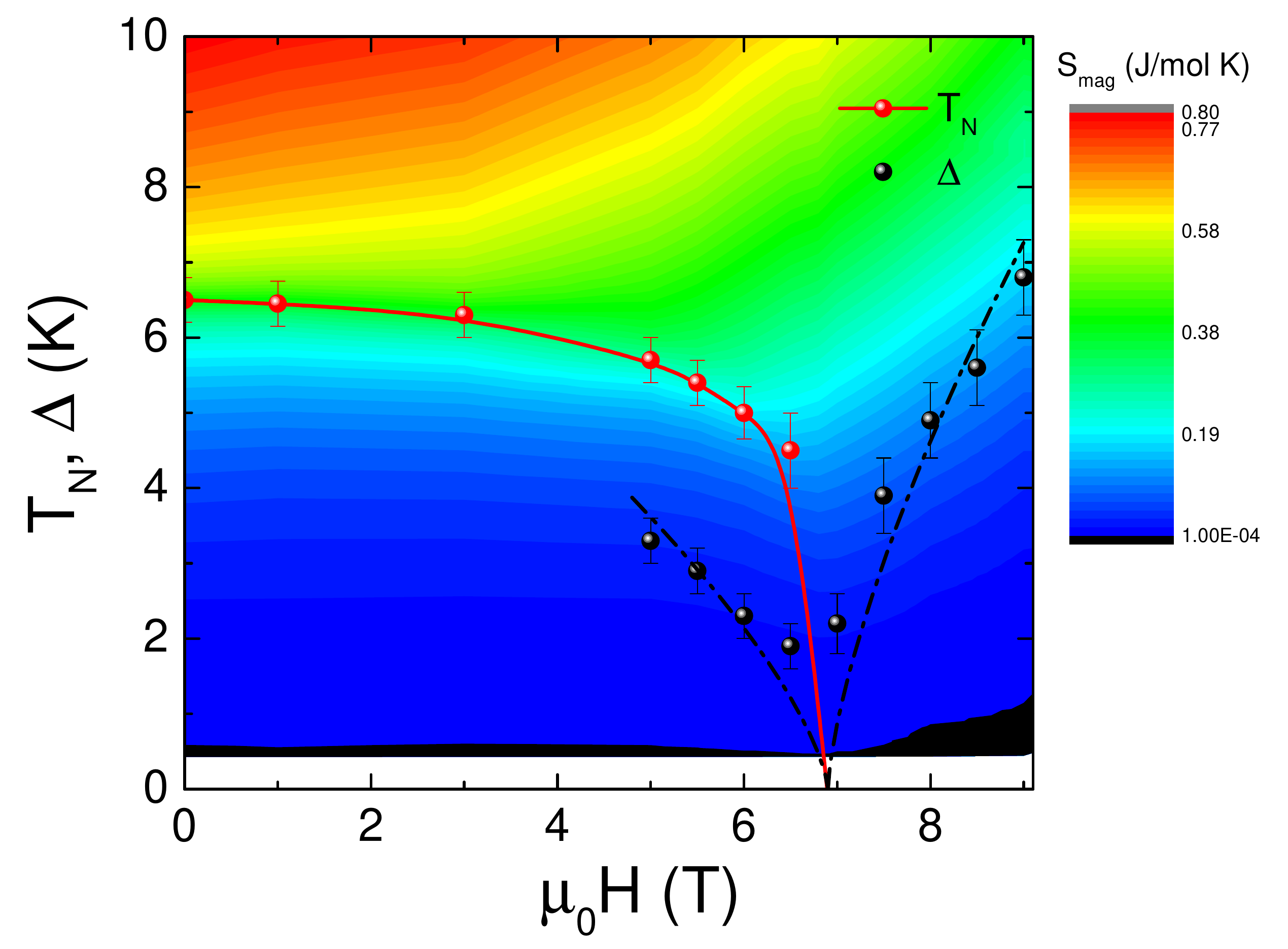}
\caption{
Phase diagram of the Kitaev magnet {\rucl} as function of in-plane field $H$ and temperature $T$, deduced from low-temperature thermodynamic measurements. Magnetic order sets in below $T_N$, with $T_N(H)$ vanishing at a critical field of $H_c\approx 7$\,T. The gap $\Delta$ also appears to vanish at $H_c$. The color code represents the magnetic entropy obtained from integrating specific-heat data.
(Figure taken from Ref.~\cite{wolter17})
}
\label{fig:rucl}
\end{figure}

The layered magnetic insulator {\rucl} is currently considered a prime candidate for realizing the physics of Kitaev's honeycomb-lattice spin liquid model \cite{plumb14,banerjee17}. At ambient conditions, the material displays antiferromagnetic order of zigzag type, with a small N\'eel temperature of $8$\,K. Microscopically, the interactions between the Ru moments are given by a combination of Kitaev, Heisenberg, and off-diagonal symmetric (so-called ``$\Gamma$'') interactions, with the Kitaev term believed to be strongest, of order $60$\,K \cite{winter16b,winter17}.

Magnetic order in {\rucl} is suppressed by the application of a moderate in-plane magnetic field of $H_c = 7$--$8$\,T. Low-temperature thermodynamic measurements \cite{wolter17}, Fig.~\ref{fig:rucl}, found approximate scaling behavior at this QPT, with exponents consistent with the Ising universality class in $(2+1)$ dimensions, in line with the low symmetry of the spin exchange. The paramagnetic phase above the critical field has been suggested to be a non-trivial spin liquid in Ref.~\cite{baek17}, but a second option is that the asymptotic high-field phase has already been reached at $H_c$ \cite{janssen17} -- this requires further study.

\subsection{Mott transition: \kaET}

The organic salts of the {\kaET} family provide a fertile ground for frustration and quantum criticality, because (i) {\ET} molecules form a triangular lattice of dimers with inherent frustration and (ii) the materials are rather soft such that their electronic properties can easily be tuned by external pressure. The latter enables access to pressure-driven insulator-to-metal transitions, such that many of the materials are located ``close'' to a Mott transition.

\begin{figure}
\center
\includegraphics[width=0.8\linewidth]{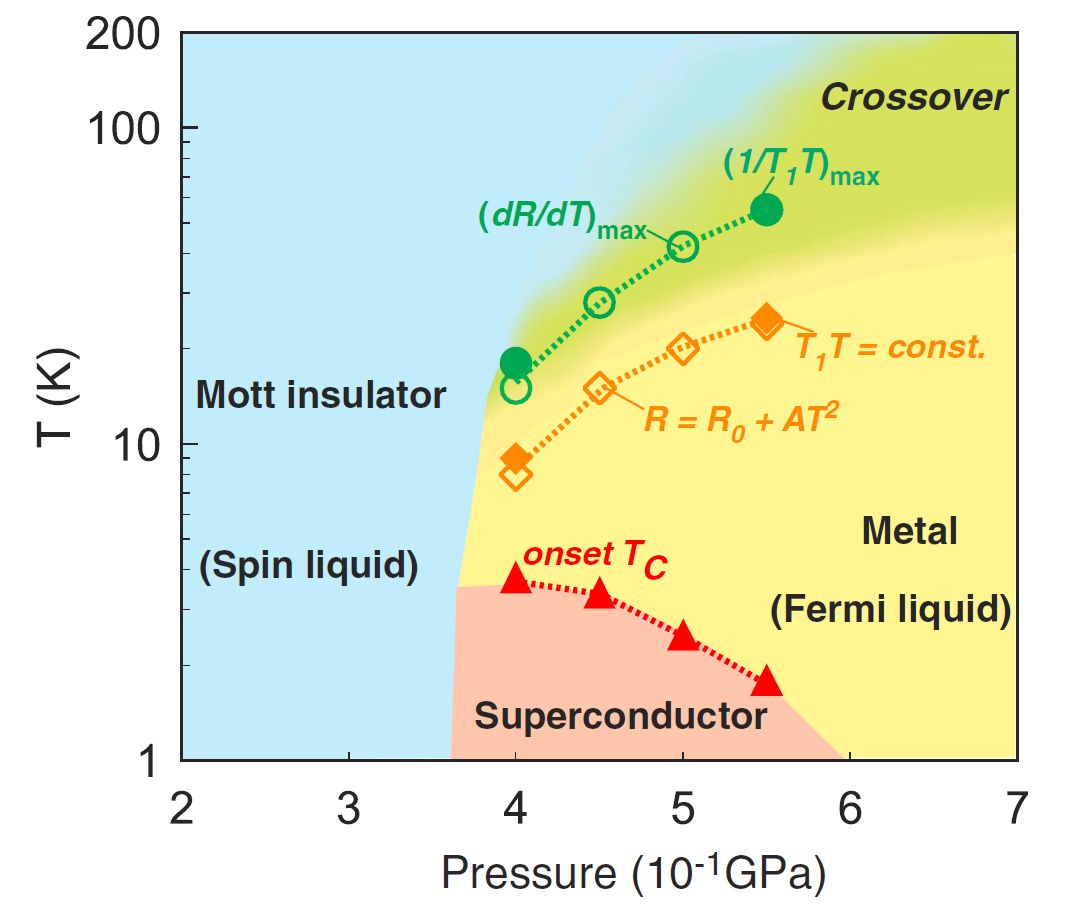}
\caption{
Temperature-pressure phase diagram of {\kaETCN}, with a pressure-driven QPT between a spin-liquid Mott insulator and an unconventional superconductor. The parent metal of the latter appears to be a conventional Fermi liquid.
(Figure taken from Ref.~\cite{kurosaki05})
}
\label{fig:et}
\end{figure}

Particularly interesting is \kaETCN: This has been characterized as a spatially anisotropic triangular-lattice spin-$1/2$ magnet, with antiferromagnetic exchange coupling of order 250\,K and no apparent order down to $30$\,mK. Hence, {\kaETCN} is a candidate for quantum spin liquid \cite{shimizu03,kurosaki05}.
Numerical investigations of the triangular-lattice Hubbard model have suggested that a spin-liquid Mott insulator may indeed occur at intermediate $U$ before the system enters the familiar 120$^\circ$-ordered state \cite{imada02,tremblay06,kawakami08}. This has been rationalized by the presence of significant further-neighbor and multi-spin exchange interactions near the Mott transition \cite{lauchli10}.
Experimentally, information concerning the presence or absence of a zero-field spin gap in {\kaETCN} is controversial. A $\mu$SR experiment \cite{pratt11} suggested the existence of two field-driven QPTs, at 5\,mT and 4\,T, respectively, with the elevated-field phases being different types of antiferromagnets. However, the interpretation of the experiments may be complicated by the effect of Dzyaloshinskii-Moriya interactions: It has been pointed out that this generates an effective staggered field which contributes to the $\mu$SR linewidth and can produce non-trivial crossover phenomena \cite{winter16}.

Interestingly, the magnetic susceptibility of {\kaETCN} extracted from torque data displays scaling as function of temperature and field, reminiscent of quantum critical behavior, in a range of temperatures up to 2\,K and fields up to 10\,T \cite{isono16}. This scaling is only cutoff at lowest temperatures and fields and suggests the existence of (hidden?) a zero-field quantum critical point, see also Sec.~\ref{sec:notuning} below.

Applying hydrostatic pressure to {\kaETCN} closes the Mott gap and drives the system metallic; this is naturally interpreted by assuming that pressure increases the electronic bandwidth and hence reduces the effective correlation strength, i.e., the ratio $U/t$ in a Hubbard-model description. The insulator-to-metal transition is found to be first order, and unconventional superconductivity appears on the metallic side at low temperatures \cite{kurosaki05}, Fig.~\ref{fig:et}.

Remarkably, {\kaET} has been successfully doped \cite{oike15}: The material $\kappa$-(ET)$_4$Hg$_{2.89}$Br$_8$, with electron doping of roughly 11\% per dimer site, can be tuned from a strongly correlated metal, viz. a doped spin liquid, to a weakly correlated metal by applying pressure. This is nicely seen in Hall-effect measurements which indicate a crossover from a small to a large Fermi surface \cite{oike15}; this suggests a fascinating interpretation in terms of a pressure-driven transition between FL$^\ast$ and FL, see Sec.~\ref{sec:dopedmott}. More detailed experimental studies are clearly called for.

\subsection{Frustrated Kondo lattice: CePdAl, YbAgGe}
\label{sec:cepdal}

A number of heavy-fermion systems, like CePdAl, CeRhSn, and YbAgGe, crystallize in the hexagonal ZrNiAl structure, with the rare-earth ions located on a frustrated lattice of equilateral corner-sharing triangles -- a distorted Kagome lattice. Signatures of local-moment frustration have been discussed for all of them; here we focus on CePdAl and YbAgGe where quantum criticality has been studied in some detail; CeRhSn will be mentioned in Sec.~\ref{sec:notuning} below.

\paragraph{CePdAl.}

In metallic CePdAl, magnetic order sets in at $\TN=2.7$\,K, with the remarkable property that three inequivalent Ce sites form, with only two exhibiting an ordered moment according to neutron diffraction \cite{doenni96}. The absence of a moment on the third site, which might be suppressed due to partial Kondo screening, has been related to frustration. The partial magnetic order can be suppressed by doping: $\TN$ vanishes at $x_c\approx0.15$ in CePd$_{1-x}$Ni$_x$Al. At $x_c$ the specific-heat coefficient diverges logarithmically, which has been interpreted in terms of 2D LGW criticality \cite{fritsch14}.

CePdAl displays rich behavior in a magnetic field \cite{lucas17}: Magnetic long-range order disappears above $4.2$\,T in a three-step fashion: The low-field antiferromagnetic phase terminates at about $3.3$\,T and is followed by two intermediate magnetic phases, with metamagnetic transitions in between. At the lowest of these transitions, the system enters a $1/3$ magnetization plateau which has been argued to increase the local-moment frustration as evidenced by an additional accumulation of low-temperature entropy. This hints at a fascinating interplay of Kondo screening and frustration which is not fully understood to date.

\paragraph{YbAgGe.}

The metamagnetic heavy-fermion metal YbAgGe orders antiferromagnetically at zero field via a first-order transition at $\TN=0.65$\,K, much below the Curie-Weiss temperature $-\TCW=15$\,K, indicating strong frustration. For applied in-plane magnetic field the phase diagram is extremely rich, with at least five different symmetry-broken phases and their transitions identified via thermodynamic and transport measurements \cite{schmiedes11}. While some of the field-driven QPTs appear to be first order, signatures of field-induced quantum critical behavior both near $4.5$\,T and near $7.2$\,T were reported in Ref.~\cite{schmiedes11}.

A detailed study of the magnetocaloric effect YbAgGe \cite{tokiwa13} suggested that the approximate singularities near $4.5$\,T to arise from quantum bicriticality: A bicritical point exists between two of the field-induced phases whose critical temperature is very small such that the behavior at elevated temperatures resembles that near a quantum bicritical point, with corresponding scaling in the magnetic Gr\"uneisen parameter. A thorough theoretical modelling is lacking to date.

\subsection{Apparent quantum criticality without fine-tuning: CeRhSn, \ybal, \priro, \irhyp}
\label{sec:notuning}

A remarkable set of experimental findings concerns apparent quantum criticality of materials without any fine-tuning of parameters. Here ``apparent quantum criticality'' refers singular thermodynamic behavior, most prominently a Gr\"uneisen parameter which diverges as a function of temperature; according to Ref.~\cite{markus} such a divergence signifies the presence of a QCP, see also Sec.~\ref{sec:qcprimer}. In the following we list a few relevant materials and then speculate about possible reasons for this unusual thermodynamic singularity; a detailed exposition of experimental data can be found in Refs.~\cite{gegenwart16,gegenwart17}.

\paragraph{CeRhSn.}

\begin{figure}
\center
\includegraphics[width=0.8\linewidth]{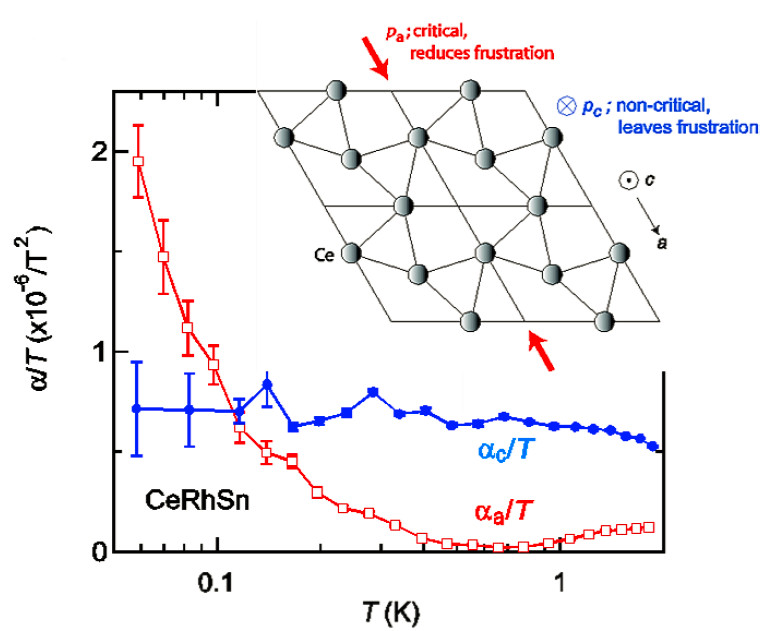}
\caption{
Thermal expansion coefficients of CeRhSn, plotted as $\alpha/T$, as function of $T$ for two different crystallographic axes. The in-plane response $\alpha/T$ diverges, while the out-of-plane response appears non-critical. The inset visualizes the distorted kagome lattice of Ce ions.
(Figure adapted from Ref.~\cite{cerhsn_grueneisen})
}
\label{fig:cerhsn}
\end{figure}

CeRhSn is a heavy-fermion metal with local moments residing on a distorted kagome lattice, stacked along the crystallographic c axis. It does not display magnetic order down to $50$\,mK \cite{cerhsn_musr}, while its specific-heat coefficient and susceptibility keep increasing until the lowest measured temperatures. The apparent specific-heat divergence is cut-off by the application of a moderate magnetic field. Both the Gr\"uneisen ratio $\Gamma$ and its magnetic counterpart $\Gamma_H$ appear to diverge at $H=0$, consistent quantum criticality at zero field and pressure \cite{cerhsn_grueneisen}. A detailed analysis shows that the divergence of $\Gamma$ originates from the in-plane contribution of the thermal expansion, whereas the c-axis contribution is non-critical, Fig.~\ref{fig:cerhsn}. This suggests a relation to frustration, as uniaxial in-plane distortions can be expected to shift the balance of frustrated interactions on the kagome lattice, while these would be unaffected by a c-axis compression \cite{cerhsn_grueneisen}.

\paragraph{\ybal.}

The intermetallic mixed-valence compound {\ybal} is a very clean metal which displays superconductivity below $80$\,mK. Above this temperature, it has been found to display a zero-field divergence of the specific heat and the magnetic Gr\"uneisen ratio $\Gamma_H$. These singularities are cut-off by an applied magnetic field; this is accompanied by a striking scaling behavior of the magnetization as function of temperature and field \cite{matsumoto11}. It has been suggested that this behavior reflects the properties of a stable non-Fermi liquid phase rather than that of a zero-field critical point, as magnetic order only appears upon applying a sizeable pressure of $2$\,GPa \cite{tomita16}.

Theoretically, a heavy-fermion model with a symmetry-dictated momentum-space vortex in the hybridization function between conduction and $f$ electrons has been proposed \cite{ramires12}. The resulting quartic dispersion gives rise to a singular density of states which, if pinned to the Fermi level, can explain much of the experimental observations. The reason for the required level pinning is unknown.

\paragraph{\priro.}
The pyrochlore iridate {\priro} is interesting for a number of reasons: First, it features Pr ions with a non-Kramers doublet ground state on the pyrochlore A sublattice, making it a candidate for ice-like physics \cite{nakatsuji06}. Second, the bandstructure involving Ir ions with strong SOC has been argued to display a quadratic band-touching point near the Fermi level. In the presence of Coulomb interactions, quadratically touching bands at Fermi level are expected to produce generic non-Fermi liquid behavior \cite{kondo15}.

Experimentally, {\priro} is metallic, albeit with a very small carrier concentration. It hence realizes a frustrated Kondo-lattice material \cite{nakatsuji06}.
Remarkable is the appearance of an anomalous Hall signal below $1.5$\,K, without signatures of magnetic ordering at this temperature \cite{machida10}; glass-like freezing of moments occurs only below 0.3\,K. Taking the anomalous Hall signal as an indication of spontaneously broken time-reversal symmetry, this has been interpreted in terms of a chiral spin-liquid state of 4f moments \cite{machida10}. Alternative interpretations, based on field-tuned frustration effects in a spin-ice-type state, have been put forward in Refs.~\cite{udagawa13,flint13,chen16}.

Equally puzzling is the observation of a divergent magnetic Gr\"uneisen ratio $\Gamma_H$ at elevated temperatures, i.e., above 0.4\,K, or at fields above 0.35\,T \cite{pr2ir2o7_grueneisen}. This suggests that the anomalous state develops from an instability of a quantum critical system.

\paragraph{\irhyp.}

The final example is an insulating magnet, \irhyp, with Ir moments on a geometrically frustrated hyperkagome lattice. The material is a weak Mott insulator and considered as a candidate for a three-dimensional quantum spin liquid, as long-range order was found to be absent down to $2$\,K, far below the exchange scale of $300$\,K \cite{okamoto07}. Subsequently, NMR measurements have detected glassy spin freezing below $\Tf=7$\,K, which may be related to the influence of quenched disorder, possibly on the Na sites \cite{shockley15}. Careful thermodynamic measurements of {\irhyp} down to millikelvin temperatures have revealed the presence of gapless excitations and a divergence of the magnetic Gr\"uneisen parameter $\Gamma_H$ in zero field. The latter has been tentatively assigned to the proximity of the material to a zero-field QCP \cite{irhyp_grueneisen}.

\paragraph{Scenarios.}
For the examples listed above, the presence of a conventional zero-field ambient-pressure quantum critical point appears unlikely, because (i) criticality without fine-tuning should be rare, (ii) a symmetry-broken phase in the immediate vicinity has not been identified.
At this point it is unclear whether there is a common mechanism behind the apparent singular behavior in the different materials. Ref.~\cite{gegenwart17} suggested distinct origins for the divergence of $\Gamma_H$: geometric frustration in CeRhSn and \priro, a strange-metal phase in \ybal, and disorder-induced local-moment effects in \irhyp.
To our knowledge, a concise theoretical picture has not been developed for any of these scenarios.

A promising direction in the context of local-moment frustration could be as follows: It is conceivable that there exists an intermediate-temperature regime with a large entropy arising from a nearly degenerate manifold of states (e.g. the ice manifold in \priro). The associated entropy is quenched by the application of a magnetic field. The effects of Kondo coupling and magnetic field may conspire as to produce a divergent magnetic Gr\"uneisen parameter. We note that such a theory also needs to explain that the divergence of $\Gamma_H$ in {\priro} is essentially independent of field direction (in contrast to strongly anisotropic field effects on classical spin ice).


\section{Outlook}

Frustrated magnetism and quantum criticality both constitute highly active fields of research in condensed-matter physics, and both have received additional fuel in the last decade by the improved understanding of topological phenomena in solids. This review article aimed to summarize the interplay of both, frustration and quantum criticality, with focus on theoretical ideas and concepts as well as links to current experiments in correlated-electron materials.

While quantum criticality in clean insulators is mainly well understood, frustration brings in new ingredients -- large degeneracies, order by disorder, and fractionalization -- which often change the rules of the game, and we have discussed a few particularly fascinating outcomes. In metallic systems, the physics of quantum phase transitions is more complicated in general, due to the presence of low-energy fermions, with many open questions even without frustration. We have highlighted different avenues to frustrated metals and pointed to intriguing quantum critical phenomena, many of which are far from understood. In addition, we have emphasized the non-trivial role played by quenched disorder in real solids as well as other ingredients such as quantum multicriticality.

We expect progress in the field to come from various directions.
First, careful measurements on new and improved frustrated materials, with specific tuning by pressure, strain, or magnetic field, will uncover novel phenomena, critical and otherwise. Those will trigger the development of new concepts.
Second, the combination of standard field-theory tools with new ideas on dualities plus the progress in the analysis of holographic models will enhance our understanding of the relevant quantum field theories.
Finally, the improvement of numerical methods, most notably matrix-product and tensor-network methods as well as sign-free Quantum Monte Carlo simulations, will enable more accurate studies of relevant microscopic models with access to critical phenomena.


\ack 

I thank E. Andrade, B. B\"uchner, S. Dey, L. Fritz, M. Garst, P. Gegenwart, L. Janssen, D. G. Joshi, V. Kataev, H. von L\"ohneysen, T. Meng, R. Moessner, S. Rachel, A. Rosch, C. R\"uegg, S. Sachdev, K. P. Schmidt, U. Seifert, T. Senthil, N. Shannon, J. van den Brink, E. Wolf, A. Wolter, P. W\"olfle, and F. Zschocke for instructive discussions and for collaborations on related work.
This research was supported by the DFG via SFB 1143 and GRK 1621.


\end{document}